\documentclass[11pt,a4paper,twoside,openright]{report}

\usepackage[utf8]{inputenc}
\usepackage[mieic,final]{feupteses}
\usepackage{graphicx,rotating,booktabs}
\usepackage{array}

\usepackage{color}
\definecolor{cloudwhite}{cmyk}{0,0,0,0.025}

\usepackage{listings}
\graphicspath{{figures/}}

\begin{document}

%%----------------------------------------
%% Information about the work
%%----------------------------------------
\title{Multimodal vs. Unimodal Physiological Control in Videogames for Enhanced Realism and Depth}
\author{Gonçalo Filipe Lopes Coelho Amaral da Silva}

%% Uncomment next line for date of submission
\thesisdate{February 7, 2014}

%%Uncomment next line for copyright text if used
%\copyrightnotice{Name of the Author, 2008}

\supervisor{Supervisors}{Rui Rodrigues (PhD), Pedro Alves Nogueira (MSc)}

%% Uncomment next line if necessary
% \supervisor{Co-Supervisor}{Pedro Alves Nogueira (MSc)}

%% Uncomment committee stuff in the final version if used
\committeetext{Approved in oral examination by the committee:}
\committeemember{Chair}{José Manuel de Magalhães Cruz (PhD)}
\committeemember{External Examiner}{Hugo Alexandre Paredes Guedes da Silva (PhD)}
\committeemember{Supervisor}{Rui Pedro Amaral Rodrigues (PhD)}
\signature

%% Specify cover logo (in folder ``figures'')
\logo{uporto-feup.pdf}

%% Uncomment next line for additional text  below the author's name (front page)
%\additionalfronttext{Preparação da Dissertação}

%%----------------------------------------
%% Preliminary materials
%%----------------------------------------

% remove unnecssary \include{} commands
\begin{Prolog}
  \chapter*{Abstract}

In the last two decades, videogames have evolved in a nearly explosive way, from text-based adventures and pixelated graphics to the near-realistic 3D environments that are common in the super-productions of big game companies (AAA titles). Although this tendency was not mirrored in the games' interaction devices area, recent studies in the Human-Computer Interaction field (HCI) have explored biofeedback interaction as an alternative to the current devices.

In the context of HCI and videogames, this dissertation explores direct biofeedback: the explicit manipulation of physiological data (i.e. one's own body) in order to perform actions inside the game. Usually, prototypes of this biofeedback type bind to each game mechanic a single physiological sensor. Although this integration is easy and quick, we consider that it has the pitfall of being too simple - it does not yet justify changing from the act of pressing a button to run to a physical action such as raising one's foot. As a consequence, this excessive simplicity wastes the vast potential that we believe to exist in this technology, to create games of increasing complexity and higher depth.

To answer to these limitations, in this dissertation we strive down an unexplored path by combining more than one sensor in a single game mechanic, which we call Multimodal Direct Biofeedback. Using the Unreal Development Kit, we created a First-Person Shooter capable of simulating reliably real-life actions, such as attenuating the recoil of a gun and underwater exploration by using the player's breathing data, among other mechanisms. To assess pros and cons of this new biofeedback type, we developed three distinct versions of the game: Vanilla (Keyboard/Mouse, for modern games), Unimodal Biofeedback (Keyboard/Mouse + 1 sensor per each game mechanic) and Multimodal Biofeedback (Keyboard/Mouse + 2 sensors per each game mechanic).

This prototype was tested by 32 participants whom compared the three versions of the game mechanisms (and each version as a whole) in terms of Fun, Ease of Use, Originality, Playability and Favourite Condition along with another questionnaire targeted at other parameters (IMI Questionnaire). Additionally, the participants compared the three versions of the mechanisms by using 12 keywords which are relevant to the game design of the game, and provided a great quantity of feedback and commentaries about the various elements of the three versions.

In compliance with previous studies, both biofeedback versions were considered more fun than the vanilla version. The unimodal and multimodal versions received similar scores in parameters such as "Fun" or "Playability", but both were appreciated by the players: the unimodal version for its simplicity of use, and the multimodal for its realism, activation safety and depth added to the game. Our biggest contribution is that multimodal biofeedback can have a quite relevant impact in terms of added depth, depending on the way it is used inside the game. On a boundary case, it can be used to increase the feeling of empowerment on the player when using certain abilities, or to intentionally make in-game actions more difficult by demanding more physical effort from the player. We believe that both biofeedback types should be combined simultaneously depending on the sensations that game designers wish to convey to the players.

\chapter*{Resumo}

Nas duas últimas décadas, a área dos jogos de vídeo evoluiu de uma forma quase explosiva a partir das aventuras em texto e gráficos pixelizados para os ambientes 3D quási-foto-realísticos comuns nas super produções das grandes empresas. Apesar desta tendência não se ter verificado nos dispositivos de interação com os jogos, estudos recentes no campo da Interação Pessoa-Máquina (IPM) têm explorado a interação através de \emph{biofeedback} como uma alternativa aos dispositivos atuais. No contexto da IPM e jogos de vídeo, esta dissertação explora unicamente a vertente do biofeedback direto: a manipulação explícita dos dados fisiológicos (ou seja, o próprio corpo) para realizar ações dentro do jogo. Os protótipos tradicionais deste tipo ligam a cada mecânica de jogo um único sensor fisiológico e embora esta integração seja fácil e rápida, consideramos que peca por ser demasiado simples e ainda não justifica a troca do ato de carregar num botão por uma ação física como levantar um calcanhar para correr. Por sua vez, esta simplicidade excessiva desperdiça o potencial vasto que acreditamos existir nesta tecnologia para criar jogos complexos e de maior profundidade.

Em resposta a estas limitações, nesta dissertação enveredámos por um caminho não explorado ao combinar mais do que um sensor em cada mecânica de jogo, a que chamamos Biofeedback Multimodal Direto. Foi criado no \emph{Unreal Development Kit} um \emph{First-Person Shooter} capaz de simular fiavelmente ações reais como conter o recuo de uma arma, levar a cabo exploração subaquática, entre outros. Para conhecermos os prós e contras deste novo tipo de biofeedback, foram desenvolvidas três versões distintas: Standard (Teclado/Rato, para os jogos atuais), Biofeedback Unimodal (Teclado/Rato + 1 sensor por mecânica de jogo) e Biofeedback Multimodal (Teclado/Rato + 2 sensores por mecânica).

Este protótipo foi testado por 32 voluntários que compararam as três variantes dos mecanismos do jogo (e cada variante na totalidade) em termos de Diversão, Facilidade de Uso, Originalidade, ``Jogabilidade'', Condição Preferida e utilizando um questionário direcionado a outros parâmetros (IMI Questionnaire). Adicionalmente, avaliaram as três variantes dos mecanismos com recurso a 12 palavras-chave relevantes ao \emph{game design} do jogo e forneceram um vasto leque de comentários sobre os vários elementos das três variantes.

Em concordância com estudos anteriores, ambas as versões de \emph{biofeedback} foram mais divertidas que do a versão standard. As variantes unimodal e multimodal receberam pontuações semelhantes em parâmetros como "Diversão" ou "Jogabilidade", mas ambas foram apreciadas pelos jogadores: a unimodal pela sua simplicidade de uso, e a multimodal pelo seu realismo, segurança de ativação e profundidade acrescentada ao jogo. O nosso maior contributo é que biofeedback multimodal pode ter um impacto bastante relevante em termos de profundidade acrescentada, dependendo do uso que lhe for dado dentro do jogo. Nos casos extremos, pode ainda servir para aumentar a sensação de poder no jogador ou para dificultar intencionalmente uma ação dentro do jogo ao exigir mais esforço físico. Acreditamos que ambos os tipos devem ser combinados em simultâneo em função das sensações que os criadores de videojogos queiram transmitir aos jogadores.
 % the abstract
  \chapter*{Acknowledgements}

Finally, the longest project of my Master's degree is over.
I was far from imagining all the hard work that would come with this project,
the first time I put my eyes on it. But even after all
the tough times and opting to work on it for six extra months,
I would still choose to work on this project all over again. More than my contribution
to biofeedback research applied to videogames, this work is
also dedicated to all the people who make or love videogames. This project
wouldn't exist without you!

To my supervisors, Professor Rui Rodrigues and Pedro Nogueira,
a huge ``thank you'' for believing that I was the man for the job,
despite my not-so-long experience in games development, and
for accepting my suggestions that took this project ``to the next level''
(in a manner of speaking).

To Chris Holden, for allowing us to use and modify his original
"Dungeon Escape" UDK map --- a simple thing as a ``yes'' made all the difference in this project
and allowed us to reach the desired graphical quality of the game.
A huge thank you!

I feel obliged to give credit to the author of \emph{``Game Design Workshop''} --- Tracy Fullerton ---,
who wrote a wonderful book on Game Design which inspired me over the course of this project. It's in the little details and additional effort of this thesis that this book's magic is present.

Finally, to those dearest to me: to my parents, family and dogs for keeping me healthy
and mentally sane during all this time; to my close friends for all the fun times, the \emph{very early} pilot testing and helping me push forward; and a special thanks to my girlfriend, as
she put it nicely, ``for putting up with me all the time''. There is no
better researcher than a happy one.

Thank you!

\vspace{10mm}
\flushleft{Gonçalo Filipe Silva}
  % the acknowledgments
  \cleardoublepage
\thispagestyle{plain}

\vspace*{8cm}

\begin{flushright}
   \textsl{``The essence of a game is rooted in its interactive nature, \\
           and there is no game without a player.''} \\
\vspace*{1.5cm}
           Laura Ermi, Frans Mäyrä
\end{flushright}
       % initial quotation if desired
  \cleardoublepage
  \pdfbookmark[0]{Table of Contents}{contents}
  \tableofcontents
  \cleardoublepage
  \pdfbookmark[0]{List of Figures}{figures}
  \listoffigures
  \cleardoublepage
  \pdfbookmark[0]{List of Tables}{tables}
  \listoftables
  \chapter*{Abbreviations}
\chaptermark{ABBREVIATIONS}

\begin{flushleft}
\begin{tabular}{l p{0.8\linewidth}}
AI      & Artificial Intelligence            \\
DLL     & Dynamic Link Library               \\
EDA     & Electrodermal Activity             \\
ECG     & Electrocardiography                \\
EEG     & Electroencelography                \\
EMG     & Electromyography                   \\
FPS     & First-Person Shooter               \\
GAZE    & Eye Gaze Tracking Sensor           \\
GLOVE   & Hand Tracking Glove Sensor         \\
GSR     & Galvanic Skin Resistance           \\
HCI     & Human Computer Interaction         \\
HR      & Heart Rate Sensor                  \\
NUI     & Natural User Interface             \\
RESP    & Respiration Sensor                 \\
SCL     & Skin Conductance Level             \\
UDK     & Unreal Development Kit             

\end{tabular}
\end{flushleft}

  % the list of abbreviations used
\end{Prolog}

%%----------------------------------------
%% Body
%%----------------------------------------
\StartBody

%% TIP: use a separate file for each chapter
\chapter{Introduction} \label{chap:intro}

\section{Context: Evolution of Videogames and Biofeedback}

Modern videogames have drastically evolved in the past two decades
from simple text-based and pixelated graphics to near-photorealistic
worlds featuring massive amounts of available content to explore and
interact with. With the introduction of these massive worlds, videogames
have evolved into complex experiences riddled with interactive and
engaging narratives, realistic gameplay mechanics and intricately
designed artificial intelligence (AI) systems, all of which aim to
close the gap with real life as much as possible. Ultimately, videogames
are played for the exclusive, highly emotional and rewarding experience
that they deliver to players~\cite{ermi2007fundamental}. At the end of 2013 we entered the
eighth generation of videogame platforms with new input devices,
but this new generation still adheres primarily to the gamepad-type
convention. Computer platforms still adhere to the keyboard/mouse
convention, as well.

The keyboard/mouse and gamepad control schemes have been the most
popular conventions for videogames and over time these devices have
evolved along with the gaming platforms. While the game platforms
themselves have evolved in terms of computational and rendering
power, the interaction devices have evolved mostly in the direction
of usability – either to accommodate different play styles or to
become more ergonomic. Heavily marketed towards hard-core gamers, the industry produced keyboards and
mice with ergonomic layouts, customizable weight, programmable keys
or gamepads in many different shapes and sizes – mostly based on the
premise that familiarity with a certain controller guarantees a good
performance in the game~\cite{brokaw2007analysis,kavakli2002usability}. 

Throughout the years game companies have tried to create new
technologies to step away from the traditional devices, with
the most recent attempts focusing on natural user interaction
(NUI) devices: the Nintendo Wiimote, the Microsoft Kinect and
the PlayStation Move. With these new natural interaction devices,
players can now hold a hero's sword in an adventure game (\emph{``The
Legend of Zelda: Skyward Sword''}, 2011), ride an inflatable boat
through a river (\emph{``Kinect Adventures''}, 2010) or move a wizard's wand
to cast spells (\emph{``Wonderbook: Book of Spells''}, 2012).

\begin{figure}
  \begin{center}
    \leavevmode
    \includegraphics[width=0.75\textwidth]{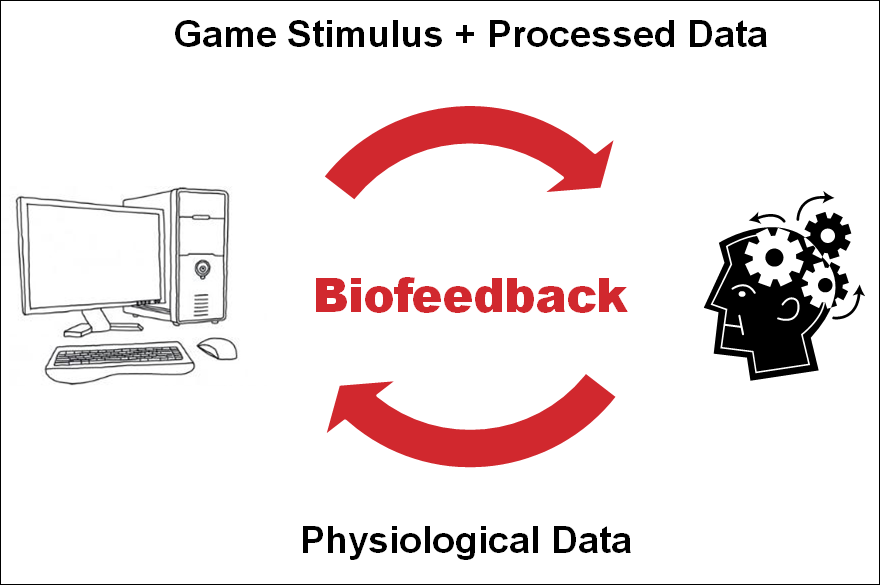}
    \caption[Biofeedback diagram - flow of physiological data]{Diagram depicting the flow of raw and processed (audiovisual) physiological data between the player and the game.}
    \label{fig:biofeed}
  \end{center}
\end{figure}

Recent research in the Human-Computer Interaction (HCI) field has
worked on reintroducing the concept of including real-time
physiological data from players in real-time applications
through the use of \textbf{Biofeedback} devices. Biofeedback was created
originally in the field of medicine in the 1980s as a training
procedure to overcome medical conditions but in the last decade
has found its way to two main applications in game HCI research
(Figure~\ref{fig:biofeed}).
These are Direct and Indirect Biofeedback:

\begin{itemize}
\item The player intentionally manipulates physiological data (i.e.
their body), which is read by the system in order to perform
actions. The player is an active agent in the transfer of
physiological data.
\item The system reads the player's physiological data and
uses it to modify internal aspects or its behaviour, but
without him/her being aware of how is physiological data
affected the game. The player is a passive agent in the
biofeedback process.
\end{itemize}

The second type is more commonly connected with \textbf{Affective Gaming}~\cite{gilleade2005affective}, where the player's emotional state is
estimated from the collected physiological data and game
parameters are modified in a meaningful way~\cite{bernhaupt2007using,
cavazza2009emotional, hjelm2000brainball, mandryk2007fuzzy, nogueira2013towards, nogueira2013automatic, nogueira2013guided, rani2005maintaining, torres2013development, yannakakis2010towards}.
It can still be considered a biofeedback system as it uses
Biofeedback instrumentation to read physiological data and
present it back to the player, hence closing an interaction
loop. This type of biofeedback system is commonly referred
to as Affective Biofeedback and is a very promising area of
Affective Computing. In this dissertation, however, we will be focusing only
on direct biofeedback.

\vspace*{12mm}

Adapting traditional videogames to make use of direct biofeedback
has been tested in the past with promising results. However, these
game adaptations have focused mostly on indirect and simple game
mechanics~\cite{nacke2011biofeedback, dekker2007please,
kuikkaniemi2010influence}. Additionally,
we believe that until now no biofeedback games have managed to
take full advantage of the physiological devices' potential.
Regarding the game design aspects, physiological sensors in these
games are integrated in a very shallow way and their contribution
to gameplay – without accounting for the different sensation of
activating an action in-game using our own body – seems uncertain
when compared to pressing a button in modern games. In~\cite{ambinder2011biofeedback},
SCL is used in \emph{``Alien Swarm''} simply to define how fast the game
timer counts down to zero. In~\cite{dekker2007please} and~\cite{kuikkaniemi2010influence},
only indirect biofeedback is explored heavily, with the player
``forfeiting game control'' (i.e. gameplay freedom) in order to
access to abilities such as invisibility, x-ray vision or faster
firing rate. Because these mechanics are activated using indirect
biofeedback, players do not know when they are going to activate
those abilities and thus fail to fully benefit from them. We consider
that biofeedback-powered abilities designed in this manner are
oversimplified and end up unbalancing the game (i.e. making it too
easy), which after a while will result in boredom due to the lack
of an appropriate challenge.

Ermi and Mäyra affirm that \emph{``the essence of a game is rooted in their
interactive nature''}~\cite{ermi2007fundamental}. Since physiological data
contains literal information on how we are controlling our own body,
we feel there is a great untapped potential in direct biofeedback games.
Given this premise, our question is whether introducing more aspects of
our physiological behaviour into the game, we can increase this
\emph{``essence of interactivity''} expressed by Ermi and Mäyra.
In our opinion, by increasing the biofeedback mechanics'
complexity we can leverage that potential to create games with far more
depth and improved game design.

\section{Document Structure}
% SECTION: OUR WORK and REPORT STRUCTURE

Before elaborating further on how we tried to increase this \emph{essence of interactivity}, we take a glimpse at the current state of the art of biofeedback research in two areas: Medicine and Games/Entertainment (Chapter~\ref{chap:sota}). Additionally, we provide a dedicated sub-section to introduce the physiological sensors used in this dissertation with examples of their previous use in research.

In Chapter~\ref{chap:requirements} we approach the requirements for our game and the criteria followed in the creation of the biofeedback game mechanics, before implementing the actual game. In Chapter~\ref{chap:chap3} we present the final game and its biofeedback game mechanics, along with the system's architecture and additional efforts done to refine the quality of our work. In Chapters~\ref{chap:chap4} and~\ref{chap:chapter5}, we present our empirical study with 32 regular players testing our biofeedback game in three interaction versions (one of them with standard keyboard/mouse interaction, and the other two augmented with biofeedback) and present their analysis of the game according to objective criteria (e.g. rating parameters of Fun on a scale of 0 to 5) and subjective criteria (associating a list of keywords to each game version and open-ended post-game commentaries).

Finally, we reflect on our work by analysing the collected data and the limitations of this study (Chapter~\ref{chap:chapter6}), presenting future avenues of research in biofeedback (Chapter~\ref{chap:chapter7}) and the conclusions of this dissertation (Chapter~\ref{chap:chapter8}).
 
\chapter{Related Work} \label{chap:sota}

\section*{}

A review of the current state of the art on biofeedback and
its closest related fields is provided in the following
sections. Most of the previous work done on physiological
interaction has focused on testing how well it applies to
medical therapy and in establishing guidelines of use for
new fields such as emotional recognition~\cite{mandryk2007fuzzy,
leon2007user, haag2004emotion, nogueira2013hybrid, nogueira2013regression}
and affective digital games ~\cite{gilleade2005affective,
yannakakis2010towards, kim2004emote}.

\section{Field of Medicine} \label{sec:medicine}

Biofeedback was first commonly used in medical therapy as a
training treatment to overcome medical conditions or patient
monitoring ~\cite{marcella1991biofeedback, blanchard1996controlled}.
In ~\cite{dong2010solution}, a mobile platform for music therapy
was presented, where users' negative emotional states were
counterbalanced through music.
In a similar approach, ~\cite{rocchi2008validation} presented a system
for body balance rehabilitation by using simple audio frequencies
to indicate correct posture.

Also in the field of rehabilitation treatments, Huang developed
an interactive biofeedback system for motor rehabilitation inside
a 3D virtual world where patients try to complete tasks while
receiving musical and visual feedback in real-time ~\cite{huang2005interactive}.
Cho, Wang and Mingyu developed EEG-based systems
that have the potential to treat medical conditions such as the
Attention Deficit Hyperactive Disorder (ADHD) ~\cite{wang2010eeg,
mingyu2006development, cho2002attention}.

Given biofeedback's seemingly easy integration with videogames and
high potential in medical treatments, various serious games have
been designed for aiding in the treatment of medical conditions,
such as: treatment of swallowing dysfunctions using EMG signals,
where a patient is a fish trying to swallow smaller fish ~\cite{stepp2011feasibility};
treatment of stuttering or similar speech disorders by reading aloud
pieces of text while monitoring the patient's GSR level
~\cite{lavender2012audition}; and General Anxiety Disorder treatment
by combining biofeedback with virtual reality systems where
certain objects inside the virtual world were modified based
on the patient's heart rate signal ~\cite{riva2010ubiquitous}.

There are also other serious games with a more ludic approach.
For example, ``Brainball'' and a relax-to-win racing game
introduce a competitive player-versus-player environment
where the most relaxed player wins the game ~\cite{hjelm2000brainball,
bersak2001intelligent}. While fun and well accomplished, they represent
a paradox in games as in a competitive environment players
feel more pressured to win and thus need to fight against
their natural instinct – they have to relax in order to
gain the advantage, which in turn generates more pressure in itself.

\section{Games and Entertainment} \label{sec:games}

Dekker and Champion modified a level in the \emph{``Half-Life 2''} game where
the Heart Rate and the Skin Conductance Level manipulated: the
avatar's movement speed and hearing ability; bullet time mechanics;
post-processing effects; dynamic weapon damage based on the sensors'
levels; invisibility and the ability to look through walls and
buildings ~\cite{dekker2007please}. While we are not sure if
the game mechanics were well balanced as a whole and avoided
making the game too easy on the players, it also featured
examples of good sensor integration such as dynamic sound effects,
reactive enemy spawner AI and the echoing of the player's heartbeat
in the game – making the game experience highly tailored and
personal for each player.

Kuikkaniemi on the other hand tried to create a more balanced
Shooter game in order to study differences between Implicit and
Explicit Biofeedback. In-game actions like walking, turning, aiming,
gun recoil intensity and firing rate were based on the player's
SCL and a RESP sensor ~\cite{kuikkaniemi2010influence}.

Rani modified dynamically the difficulty level of a Pong game
based on the user's anxiety state, which was estimated by real-time
analysis of the captured physiological data ~\cite{rani2005maintaining}.
Parnandi developed a car-racing game where factors that influenced
the game difficulty such as car speed, road visibility due to fog
and steering jitter were modified based on the arousal levels of
the player (determined by an EDA sensor) ~\cite{parnandi2013control}.

Ambinder and Valve have also made some experiments with biofeedback
on some of their games ~\cite{ambinder2011biofeedback}. The first one used the player's
arousal state via a SCL sensor to modify the AI director of the game
\emph{``Left 4 Dead''}, which modified in-game events such as enemy spawns,
health and weapon placement, boss appearances, etc. The second
experiment was performed on \emph{``Alien Swarm''} and connected the player's
arousal state to an in-game timer, where higher arousal levels made
the timer tick faster. The last experiment used eye-gaze in the game
\emph{``Portal''} by using the mouse to control the player's camera and the
eye-tracking for aiming tasks.

Nacke worked on the concept of Direct and Indirect Biofeedback to
augment interaction with a game ~\cite{nacke2011biofeedback}. The developed
prototype was a 2D side-scrolling shooter which used physiological
sensors to control the avatar's speed and jump power, the enemies'
size, the flamethrower's flame length and the weather conditions
which affected the boss of the game. Additionally, the player's GAZE
was used to paralyze enemies for a limited amount of time. Torres
created a procedurally generated horror game --- \emph{``VANISH''} ---,
which uses indirect
biofeedback to influence the game's level layout, item and enemy
spawns, enemy AI and player character abilities ~\cite{torres2013development}.

The use of GAZE in video games has been previously explored,
although not exactly in the scope of biofeedback research as Nacke
and Ambinder did. Isokoski and Martin presented a preliminary study
with a small prototype to compare the efficiency of eye-trackers
versus traditional game controllers – it used two types of aiming
simultaneously, one being the typical crosshair controlled by the
mouse and the other moved by the eye-gaze of the player ~\cite{isokoski2006eye}.
Smith and Graham ~\cite{smith2006use} tested GAZE in three different game genres:

\begin{enumerate}
\item Controlling view orientation in a \emph{``Quake 2''} clone, where
looking at an object that was not in the center of the screen
would automatically rotate the camera to it.
\item Issuing commands to a virtual avatar in the game \emph{``Neverwinter Nights''}
by first looking at the target (a point in space or a treasure chest) and
then using the left mouse button.
\item A modified version of the arcade game \emph{``Missile Command''} where the
eyes were used to aim towards missiles before pressing the fire button.
\end{enumerate}

Istance tested a modified version of the online multiplayer game
\emph{``World of Warcraft''} where all interaction with the game was done only
using GAZE dwelling ~\cite{istance2009your}.

\vspace*{12mm}

Using speech recognition and analysing facial features can be
considered a form of Biofeedback – we are still discussing
physiological features unique to people, and the player can use
that kind of information to consciously manipulate voice and facial
expressions. In Emotional Flowers, the user's facial expressions are
used to influence the growth of a virtual flower at several times during
the day ~\cite{bernhaupt2007using}. Cavazza created an interactive
storytelling system to speak with virtual characters using emotional
speech recognition capabilities: that is, the user's affect was
conveyed by his/her voice and the virtual characters would react
differently based on the user's answers ~\cite{cavazza2009emotional}.
Depending on the different responses given by the user, the narrative
of the story would be directed towards different endings. Kim created
a virtual snail that reacts dynamically to the user's emotional state,
captured by the user's voice, ECG, SCL, RESP and EMG sensors
~\cite{kim2004emote}.

Biofeedback has also been used to modify the behaviour of amusement rides:
Marshall built a bucking bronco system where the user's breathing
can manipulate directly the movements of the bronco – in a similar way
to direct biofeedback; or where the system monitored the breathing
information and changed strategies mimicking the behaviour of a human
operator – closely related to the indirect biofeedback flavour ~\cite{marshall2011breath}.
Breath control provided an interesting balance in terms of
semi-voluntary control and game design itself – as in one
of the conditions the winning strategy of holding breath had
the perk of a bronco with reduced movement, the average person
eventually has to resume breathing – thus receiving the penalty
of harder bronco movements.

Biofeedback has also been used in an automatic bookmarking system suitable
for audio books, which was able to detect when users were interrupted by
people or any real world events ~\cite{pan2011now}. In the context of
videogames, this could be used to automatically pause the game when the
player is distracted or forced to look away from the screen. Unexpected
interruptions sometimes provoke frustration on players.

\section{Primer of Used Physiological Measures} \label{sec:primer}

For a quick synopsis of physiological measures in games, see
~\cite{ambinder2011biofeedback} or ~\cite{kivikangas2010review}.

\vspace*{10mm}

\emph{Electromyography} (EMG) measures the electrical activity
of muscle tissue. There are records of EMG being used for emotion
detection ~\cite{nakasone2005emotion} and game profiling ~\cite{nacke2008flow},
but recent works have used it as a direct form of input for games
~\cite{nacke2011biofeedback}. In our study, we use it on an individual's
arms and legs for direct input into the game.

\vspace*{6mm}

A \emph{Torsion Glove} (GLOVE) is a tracker glove with embedded sensors which
can detect the degree to which each of the wearer's fingers are
individually bent or straightened, and thus falls under the category
of direct input\footnote{While this is not traditionally used in Biofeedback
research and is more resembling of Performance Capture Animation techniques,
participants in our study seem to have considered it as an attractive device
and could be used more often in research.}. In our study, we use it as a tool
to detect different hand poses. We did not find previous works using a
physical glove to track gestures. A similar alternative is to use computer vision
techniques to indirectly infer hand gestures ~\cite{bretzner2002hand, chen2007real},
although they are at a disadvantage as the glove devices can track
precise data of the wearer's fingers.

\vspace*{6mm}

A \emph{Respiration sensor} (RESP) is stretched across an individual's chest
and can be used to infer the wearer's chest volume or breathing rate
~\cite{kuikkaniemi2010influence}. Some possible game mechanics that can be derived from
this information are modulating a character's stamina exhaustion based
on the breathing rate, or using the first derivative to detect
``silent breathing'' periods in a stealth game to avoid being detected
by the enemy. In our study, we use it to infer an individual's chest volume.

\vspace*{6mm}

A \emph{Temperature sensor} (TEMP) is traditionally used to measure body
temperature, but in our study we use it by blowing hot air on it ~\cite{nacke2011biofeedback}.

\chapter{Prelude: Requirements and Design Choices} \label{chap:requirements}

\section{State of the Art Reflection}

Our analysis of the current state of the art on biofeedback games brought us to realise that, on most occasions, a game mechanic was matched to a single physiological sensor. In direct biofeedback mechanics specifically, there were no previous studies examining the combination of two or more sensors to produce a fun game mechanic~\cite{nacke2011biofeedback}. This observation provided a simple but critical aspect where we could contribute to biofeedback research in videogames. To this combination of two or more direct physiological sensors we call \textbf{Multimodal Direct Biofeedback}, and \textbf{Unimodal Direct Biofeedback} on the case where a single sensor is used. Considering this new type of biofeedback which provided a fresh perspective for game design, and the direct physiological sensors available for use, we strived to design a game focused on a meaningful use of the sensors in the game.

\section{Designing for Biofeedback Interaction}

According to Brown~\cite{brown2004grounded} and Ermi~\cite{ermi2007fundamental}, two of the aspects that define the player's Immersion in a videogame are game controls and the cognitive and physical (e.g. hand-eye coordination) challenges that derive from gameplay – in our case, the mastery of the biofeedback controls. Particularly, Brown and Cairns~\cite{brown2004grounded} state that game controls are one of the first barriers of the ``Engagement'' stage in Immersion. Thus, we can safely assume that if the controls lack in quality, players will most likely reject them. This was an aspect that we wanted to avoid in the biofeedback variants of our game.

To this end, the results collected by Nacke in his biofeedback 2D platformer provided a good starting point in the design of our prototype:

\begin{quote}

\emph{``Participants preferred physiological sensors that were directly controlled because of the visible responsiveness. (...) Physiological controls worked most effectively and were most enjoyable when they were appropriately mapped to game mechanics. (...) For example, when breathing out triggered a longer flame of the flamethrower, blowing hot air on the temperature sensor decreased the amount of snow, or flexing the leg muscle increased speed and jump height.''}~\cite{nacke2011biofeedback}

\end{quote}

However, he also mentions the added tension of realistic or natural mappings versus the highly imaginative possible scenarios of videogames~\cite{nacke2011biofeedback}: \emph{``Natural mappings may present more intuitive game interfaces, but also limit the flexibility and generality of the sensors for game control.''} 

Rather than prioritizing the game development as a standard Mouse/Keyboard game and then trying to retrofit the physiological sensors to its gameplay mechanics, we strived to design the game from the ground up based on the sensors. This was done to improve the overall game design and ultimately make the experience feel more natural – that is, to minimize players' rejection of the sensors for feeling ``forced'' or ``gimmicky'', which are recurring complaints on the more widely-commercialized natural user interface (NUI) systems~\cite{teofilo2013gemini, suma2011faast}. In compliance with the immersion theories presented above, it is of general consensus among the gamer community that if a game is unplayable due to terrible game logic or bad controls it won't be played – this was an aspect of considerable importance to us.

\vspace*{12mm}

Based on these results, we opted to use multimodal biofeedback to simulate interaction with natural mappings, but also with a more physical component in the game\footnote{This is not meant in the way of \emph{fitness games} that have been popularized by the Wii, for example. By physical component, we mean to literally describe a realistic interaction with the game.}. A great example of this is the Sprinting mechanism used in Nacke's platformer~\cite{nacke2011biofeedback}, that we decided to re-use in our game. At the same time, we maintained the possibility of using these sensors to \emph{simulate} supernatural actions that are typical in videogames or fiction works in movies and books, while keeping them based on physical actions. These, along with Nacke's findings presented above and our intention to make the physiological sensors the ``protagonists of the game'' (i.e. a central core to gameplay), were the selected design guidelines that went into the creation of the prototype.

This initial analysis led to the planning of an initial set of game mechanics to be developed using multimodal biofeedback. For example, the following mechanisms were on the initial list: attenuating the recoil of a gun through the combined use of arm action and increased chest volume, telekinesis using the GLOVE device and EMG sensors, a stealth mode combining EMG cables on the legs and breathing data, or the power to confuse enemies (this one being a special power). Some of these ideas did not survive the design and implementation phases, as they were hard to implement due to time constraints or depended directly on our choice of game engine. Other concepts, such as the stealth mode or the telekinesis, were adapted into different game mechanics that could be incorporated in the game without spending additional resources. While the evolution of the original game design over time is perfectly common in the development of consumer-level videogames, we will not discuss this process in detail as it is not in the scope of this dissertation. The final design of the game (including the biofeedback mechanisms) is presented in Chapter~\ref{chap:chap3}.

\section{Choosing a Game Engine: Time and Resources}

Creating a game from scratch is a task that can range from mere months to years of development. Although the biofeedback interaction mechanisms were meant to be created from scratch, creating other assets relevant to the game (e.g. 3D models, audio, level design, etc.) would not be possible as it would require more time and resources that were not in our reach. However, it was still important to provide a game experience similar to the one present in existing consumer-level games.

Choosing the correct engine proved to be a critical aspect to tackle both issues. To cut down the development time, we first narrowed our choices of game engines to the \emph{CryEngine SDK}, \emph{Source SDK} and the \emph{Unreal Development Kit (UDK)}. Besides the fact that they are free for non-commercial uses and are popular choices in the game industry, they ship with a small amount of assets with good quality, ready for use - which provided us the chance to design our own custom level, if necessary. A more thorough investigation of these alternatives led us to choose UDK, due to the following factors:

\begin{itemize}

\item The global architecture of the engine and the set of game classes built on top of their scripting system, \emph{UnrealScript}, was easy to understand on a first sight. Most game objects in the engine can be extended through inheritance (like in ordinary object-oriented languages), and new game features can be inserted without breaking the functionality of existing classes.

\item It provided ways to establish communication with the physiological devices (which are located outside the game engine) in an almost trivial manner, using a UDK feature called \emph{DLLBind}. As stated in the engine's documentation\footnote{UDK's \emph{DLLBind} documentation and examples - http://udn.epicgames.com/Three/DLLBind.html}, developers can create a custom dynamic-link library (DLL) written in C++ with exported functions that can be called in \emph{UnrealScript}. This allows us to inject additional features in the game that cannot be delivered by the engine itself, as is the case with our physiological devices.

\end{itemize}

This paved the way for our prototype, the ``GenericShooter 3000'', which has the basic premise of delivering to players a First-Person Shooter (built in \emph{UDK}) where they can execute real-life actions in the game using the complexity of multimodal biofeedback. We present in the next section a final version of our prototype, along with how biofeedback interaction was integrated in the game.
\chapter{Implementation: A Direct Biofeedback Game}\label{chap:chap3}

\section*{}

To compare how both multimodal and unimodal biofeedback-augmented videogames impact game design and user experience in modern First-Person Shooter videogames, we are interested in answering the following questions:

\begin{enumerate}
\item What are the players' responses to physiologically augmented
gameplay mechanics in modern First-Person Shooter games?
\item How do Multimodal and Unimodal mechanisms compare to each
other in terms of in-game actions, situation context and
player experience?
\item Can we enhance game realism and depth using Multimodal
and Unimodal Biofeedback? If so, how far?
\end{enumerate}

\vspace*{12mm}

As a means to answer the previous questions, three identical versions of the game with different interaction methods were developed:

\begin{itemize}
\item \textbf{Normal (or Vanilla)} --- Mouse and keyboard only.
\item \textbf{Unimodal} --- Mouse/Keyboard plus one physiological
sensor per augmented gameplay mechanic.
\item \textbf{Multimodal} --- Mouse/Keyboard plus two physiological
sensors per augmented gameplay mechanic.
\end{itemize}

\section{``GenericShooter 3000'' Game Description} \label{sec:gamedescritpion}

In our game the player has three guns at his/her disposal: a Physics Gun
to interact with physics-enabled objects and remove obstacles out of
the way when necessary; a Link Gun and a Shock Rifle from the
\emph{``Unreal Tournament''} games which function similarly to a machine gun
and an ordinary rifle, respectively.

Besides basic locomotion, the main character can also perform other
typical actions such as: sprint, jump, crouch, put out torches to
solve puzzles and interact with buttons/handles on walls and floors.
Lastly, there are two special abilities:

\begin{enumerate}
\item Possession --- The player enters the body of an enemy and
controls it in a first-person view. This leaves the main character's body
immobile and vulnerable to nearby enemies for the duration of the
possession. The player can use the possessed body to activate doors
and switches or to fight other enemies – on certain key points, this
ability is the only way to progress in the game. The possession technique
has no time limit and ends when the possessed body dies (e.g. fights versus
other enemies, drowning, suicidal), the player stops the ability or when
the main character's body is killed (in this case the player loses the game).

\item Invisibility --- The player becomes invisible to all enemies
and may freely roam the level. This technique ends whenever the player
interrupts it or after its power level (indicated on the lower left
corner of the screen) ends.

\end{enumerate}

Enemies run the same AI algorithm in the entire game and have either
the Link Gun or the Shock Rifle. There are no boss encounters at this
current stage. For our test scenario, the game takes place in a flooded
prison where the main character is incarcerated and has to escape.
To this end he/she must find three keys to exit the prison, which for our
test purposes is
the ``winning condition'' - that is, players have to keep playing until they
manage to escape the prison. The level is based on an existing UDK map called
``Dungeon Escape''\footnote{Created by Chris Holden, who gently authorized us
to use the map on our project. The original map is available at
http://chrisholden.net/07.htm .}, which we modified as necessary
to change the level's interaction logic (e.g. behaviour of gates, handles,
switches, spawning enemies) and to include an underwater section that did
not exist in the original level. Enemies and obstacles were placed to
force usage of all the main character's abilities at some point in the level.

\section{Game Mechanics Subject to Biofeedback Control} \label{mechanicsdescription}

There are a total of eight game mechanisms that are activated
differently depending on the game version being played.

\subsection{Gun Recoil Attenuation}
Based on popular FPS titles such as \emph{``Counter-Strike''} or
\emph{``Call of Duty''}, the Link Gun and the Shock Rifle move the
player's view whenever a shot is fired. In the vanilla
version, no attenuation is available and the recoil values
are approximated as faithfully as possible to the average
recoil exhibited in existing FPS titles.

The real-world logic behind this mechanism is that when someone
fires a gun, contracting the firing arm and sustaining the
breathing cycle will result in steadier aim and reduced
weapon kickback effects – in reality this does not necessarily
apply, but common popular wisdom dictates otherwise and thus
makes the mechanism \emph{feel real} in this way.

In the unimodal version, players can significantly attenuate
recoil by contracting their right arm (EMG-Arm). In the
multimodal version, players must both contract their right
arm (EMG-Arm) and increase their chest volume (RESP) to
apply the effect to the same intensity. Activating the
mechanism only partially (by either contracting their arm
or increasing their chest volume) results in a partial
reduction effect. Both biofeedback versions have their recoil
values slightly increased to force the usage of the
physiological sensors while at the same time trying to
approximate the behaviour of a real-life gun.

\subsection{Invisibility}

\begin{figure}
  \begin{center}
    \leavevmode
    \includegraphics[width=1.00\textwidth]{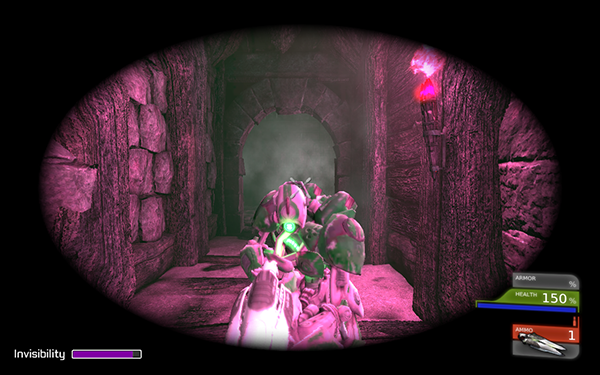}
    \caption[\emph{Invisibility} game action snapshot]{Invisibility view mode with an enemy in front. Invisibility's post-processing effects occlude the screen borders and change the color scheme.}
    \label{fig:invis}
  \end{center}
\end{figure}

The time Invisibility can remain active is the same
for the three versions – a duration bar is displayed on the
lower left corner of the screen (Figure~\ref{fig:invis}).

As there is no real-world analogue to becoming invisible,
we leveraged a common player behaviour in stealth games - becoming
tense and having very shallow, slow breathing (perhaps in a subconscious
attempt at not making noise in-game, although this behaviour of making
too much noise or breathing loudly using the sensors was not implemented in the game).
However, to balance the game's design we decided to treat it as
a special ability (based on popular fiction works such as
\emph{``Dragon Ball''}, \emph{``Eragon''} or \emph{``Harry Potter''}) which requires a
feeling of empowerment before activating powerful abilities.

In the vanilla version, players press the ``Q'' key to become
invisible. In the unimodal version, players are required to
breathe in (RESP) to activate the power, and hold their
breath in order to maintain invisibility active. If they
breathe out, the power is deactivated. In the multimodal
version, players must first make a closed fist pose
(GLOVE) - mimicking the magic seal for a spell or
to concentrate power - and then breathe in (RESP) to
activate the power, subject to the same no breathing
constraints as in the unimodal version.

\subsection{Underwater Breathing}

\begin{figure}
  \begin{center}
    \leavevmode
    \includegraphics[width=1.00\textwidth]{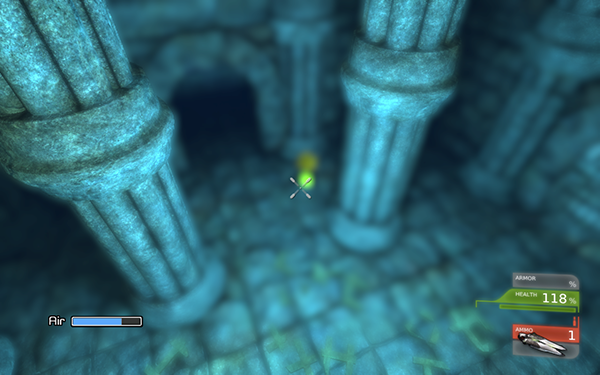}
    \caption[\emph{Underwater} post-processing effects]{Underwater post-processing effects with blue tinting and distance-based blurring.}
    \label{fig:underw}
  \end{center}
\end{figure}

Similarly to what is done in other titles such as the
old \emph{``Tomb Raider''} games, the player has limited time
to stay underwater and have their oxygen level
displayed on the screen (Figure~\ref{fig:underw}).

In the vanilla version, oxygen decreases linearly over
time without any intervention from the player.

In both biofeedback conditions the real-world analogies
are obvious: players are required to physically hold their
breath using only the RESP sensor while they are underwater.
Trying to breathe in while underwater results on filling
the lungs with water and drown instantly as a result.
While this is not what would normally happen in
real-life - that is, people would fill their lungs
with water and feel a strong need to come to the surface –,
earlier playtesting showed that players continued diving
ignoring health losses in the character. Ignoring the
RESP sensor and health losses was a behaviour that we
did not desire in the game, thus we took a more punitive
approach to this mechanism.

RESP levels while diving are also connected to the on-screen
oxygen bar (Figure~\ref{fig:underw}), replacing the vanilla version's
decreasing oxygen behaviour.

\subsection{Possession}

\begin{figure}
  \begin{center}
    \leavevmode
    \includegraphics[width=1.00\textwidth]{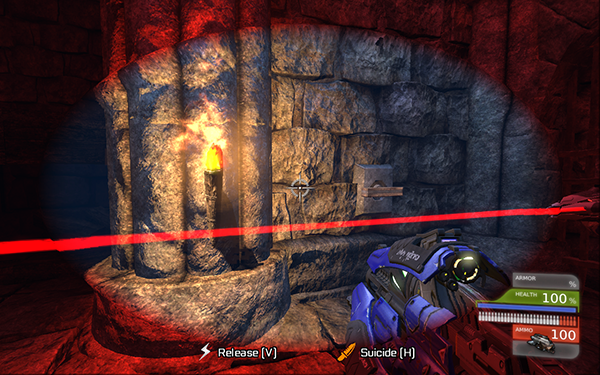}
    \caption[\emph{Possession} post-processing effects (enemy view)]{Possession post-process effects include border overlays with red color (meaning ``danger'', to be associated with enemies). Keybinds are also communicated on the bottom of the screen (on the Vanilla version only!) as a good design practice followed by videogames.}
    \label{fig:poss}
  \end{center}
\end{figure}

In the vanilla version, players take over the enemy's body
by pressing the ``V'' key (Figure~\ref{fig:poss}). Similar to other
popular titles (in this case \emph{``Dishonored''}), to return to
their original body, they can perform a Release action
using the ``V'' key, or a Suicide action using the ``H'' key.

\begin{figure}
  \begin{center}
    \leavevmode
    \includegraphics[width=0.50\textwidth]{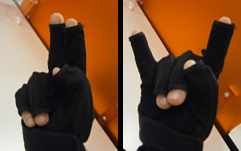}
    \caption[Biofeedback hand gestures of \emph{Possession}]{Possession hand gestures for both biofeedback versions: initiate and reverse (left), suicide (right).}
    \label{fig:possgest}
  \end{center}
\end{figure}

In the unimodal version, players Possess enemies by blowing
hot air on the TEMP sensor, which mimics the player's soul
leaving the body as it enters the possession's target. In the
multimodal version, players first perform the hand pose
displayed (Figure~\ref{fig:possgest} on the left) and then blow hot air onto
the TEMP sensor. To maintain consistency with the Invisibility
technique, this pose resembles the magic seal for a magical ability.

For both versions, to return to the original body they have
to perform one of the hand poses (reverse or suicide)
displayed on Figure~\ref{fig:possgest}.

\subsection{Fire Blow: Interaction with Fire Objects} \label{ssec:fireblow}

This action was designed to be used as a puzzle solving
ability inside the game involving torches or other fire
objects. In future work we would like for it to also be a
stealth mechanic, where blown-out torches decrease the
visibility of an AI enemy - it allows for more strategic
thinking from the player.

\begin{figure}
  \begin{center}
    \leavevmode
    \includegraphics[width=1.00\textwidth]{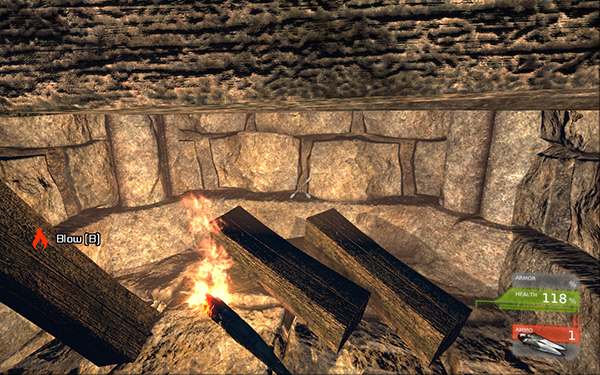}
    \caption[Example of a \emph{Fire Blow} puzzle]{A puzzle where the player has to light up the
             fireplace combining a torch and ``Fire Blow'' to feed power to a non-working gate.}
    \label{fig:fblow}
  \end{center}
\end{figure}

In the vanilla version, players put out a torch or light up a
fireplace by pressing the ``B'' button (Figure~\ref{fig:fblow}). In the unimodal
version, players blow on the TEMP sensor.

In the multimodal version players must first inhale heavily to
increase their chest volume through the RESP sensor and finish
the action by blowing on the TEMP sensor. Since the in-game
puzzles were in the size of torches or fireplaces, they
required a larger breathing effort – the multimodal version
portrays this interaction more realistically.

\subsection{Sprinting}

\begin{figure}
  \begin{center}
    \leavevmode
    \includegraphics[width=1.00\textwidth]{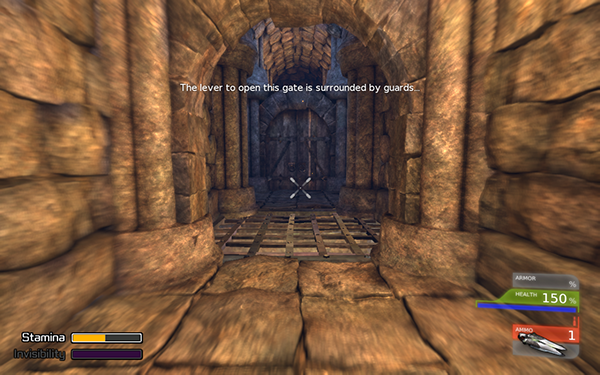}
    \caption[\emph{Sprinting} view and post-processing effects]{The Stamina bar shows on the screen when the player presses Shift or the EMG-Leg sensors are used. A blur effect is applied to the screen to enhance the sensation of running – this is a feature that already exists in the UDK engine which we left unchanged.}
    \label{fig:sprint}
  \end{center}
\end{figure}

On the lower corner of the screen is a stamina bar that
shows for how much longer the player can run before being
forced to recover his strength (Figure~\ref{fig:sprint}). Stamina decrease
over time and maximum speed are the same on the three
conditions.

In the vanilla version, Sprinting is activated by holding
down the ``Left Shift'' key. In the unimodal version, players
can lift the left leg's heel or foot tip (EMG-Leg). In the
multimodal version, players are required to use both
feet (2x EMG-Leg).

\subsection{Item Use: Using Objects or Equipping Them}

This ability is used to pick up objects inside the
game, to open locked doors or to interact with other
``usable'' game objects. In the vanilla version, players
use the ``E'' key.

\begin{figure}
  \begin{center}
    \leavevmode
    \includegraphics[width=1.00\textwidth]{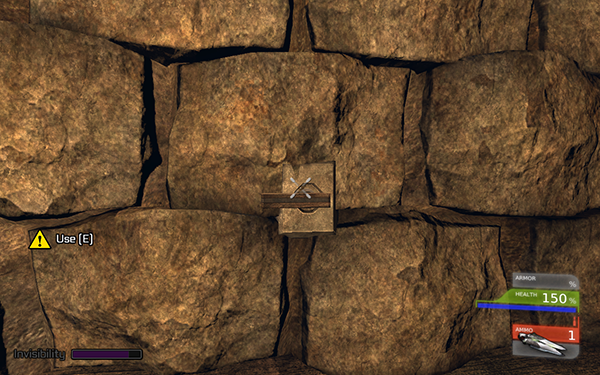}
    \caption[Example of an \emph{Item Use} action]{``Item Use'' works on handles to open certain doors,              other interactive objects or to equip key items like the Prison Keys (which are required to finish the level).}
    \label{fig:use}
  \end{center}
\end{figure}

In the unimodal version, players close the hand equipped
with the GLOVE device (Figure~\ref{fig:use}). In the multimodal
version, players are required to contract the right
arm (or alternatively to move it in any direction)
while grabbing the same object with the GLOVE.

\subsection{Grabbing Objects (Physics Gun)}

To use this action players have to equip the Physics Gun.
In the vanilla version, this is activated by dragging the
Left Mouse Button. In the unimodal version, players close
their hand (GLOVE) to grab the object and open it to
release the object - to move it while grabbing from one
place to another the player can use the WASD keys to move
or use the mouse to rotate the camera.

Unfortunately, the GLOVE device did not have an accelerometer
and this was an ability that we wanted to perform purely on
the GLOVE – players who wished to rotate the camera were
forced to do so by moving the mouse with the tip of their
closed hand. This is a usability aspect that we wish to
improve in future work.

\begin{figure}
  \begin{center}
    \leavevmode
    \includegraphics[width=1.00\textwidth]{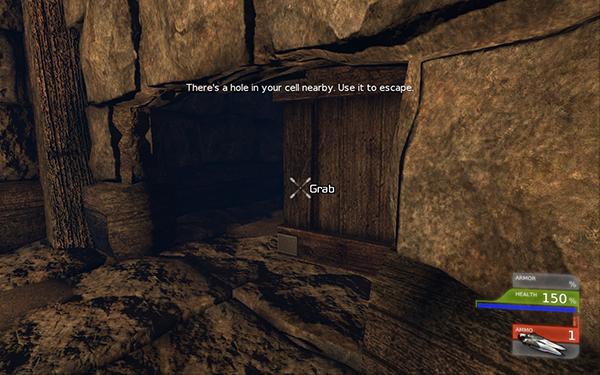}
    \caption[Example of a \emph{Grab Object} obstacle]{A physics-enabled object is covering a hole where the player has to pass through by clearing the way with the Physics Gun.}
    \label{fig:grab}
  \end{center}
\end{figure}

In the multimodal version, players first contract the
right arm (EMG-Arm) as in the ``Item Use'' ability until
the ``Grab'' word on the screen (Figure~\ref{fig:grab}) is lit green,
and then close their hand (GLOVE) to grab the object - at
this stage the arm can be relaxed as the object is being
held successfully.

\section{System Architecture} \label{sec:architecture}

\begin{figure}
  \begin{center}
    \leavevmode
    \includegraphics[width=0.50\textwidth]{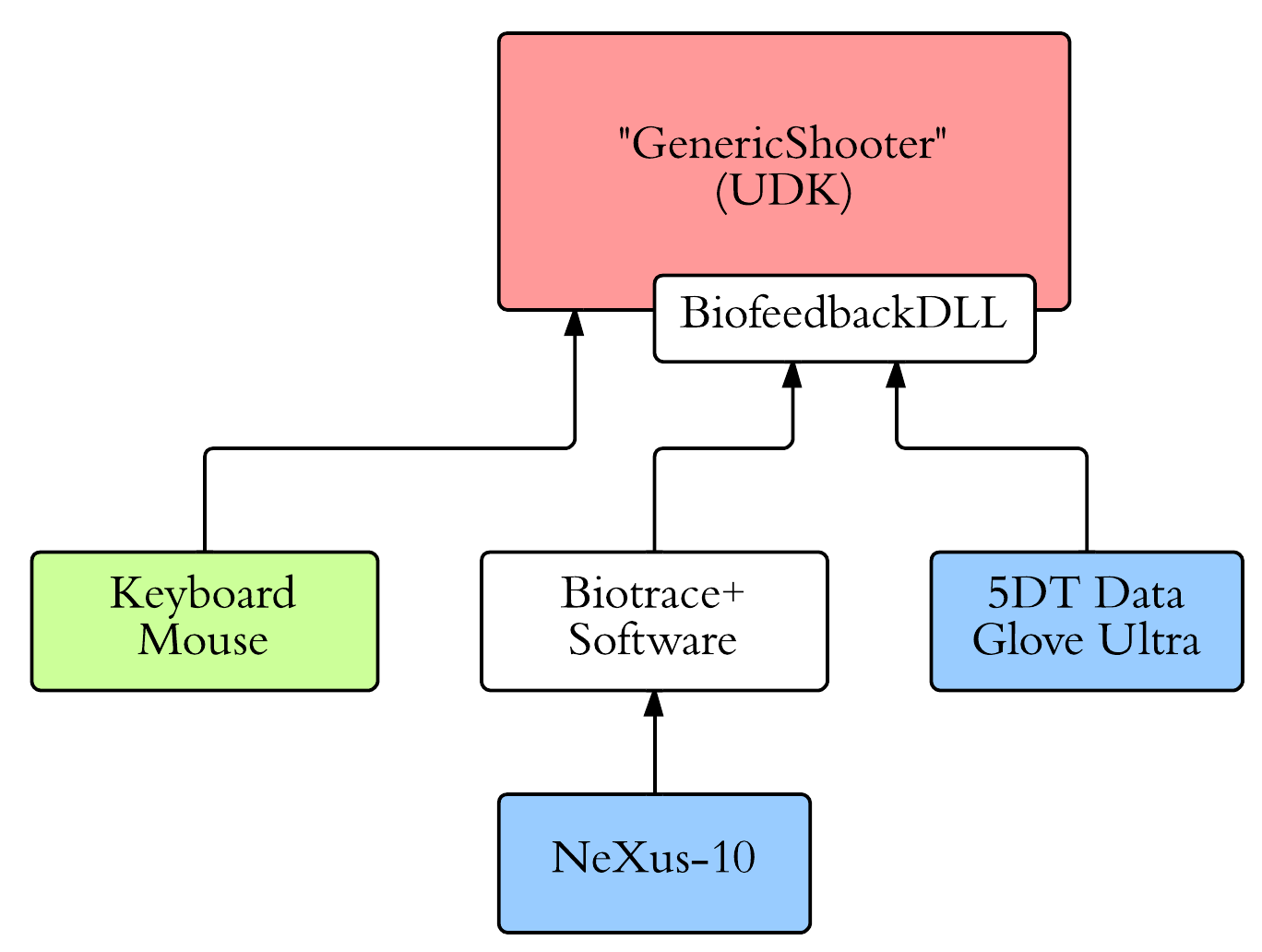}
    \caption[System architecture for the biofeedback versions]{System architecture in the biofeedback versions of the game. Players see the result of their actions in the UDK engine.}
    \label{fig:architecture}
  \end{center}
\end{figure}

To create additional functionality, UDK has the ability
to integrate custom developed Windows Dynamic-link Libraries
(DLLs) with functions that can be called from inside the
UDK's scripting language (UnrealScript). We wrote a custom DLL
in C++ which is responsible for retrieving physiological data
from outside UDK, resulting in the overall system architecture
depicted on Figure~\ref{fig:architecture}.

The \emph{NeXus-10} device is responsible for capturing physiological
data for the EMG, RESP and TEMP signals. Signal processing here
is handled by the \emph{Biotrace+} software suit, which also performs
the signal acquisition from the NeXus-10 device via Bluetooth.
The GLOVE's data stream is accessed by a C++ library provided
by the glove's manufacturer.

The \emph{BiofeedbackDLL} provides normalized data into the the game
after it is processed by the \emph{Biotrace+} Software suite, as we
describe now in more detail.

\section{Calibration of Physiological Devices}

\begin{figure}
  \begin{center}
    \leavevmode
    \includegraphics[width=1.00\textwidth]{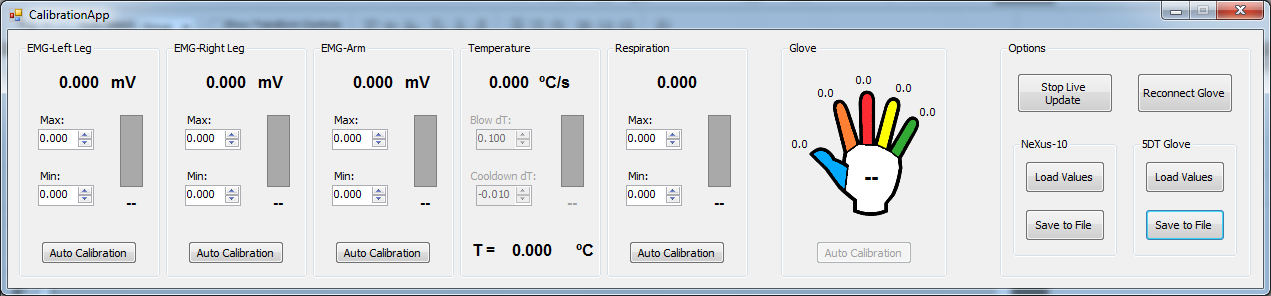}
    \caption[\emph{CalibrationApp} tool snapshot]{Snapshot of the application for calibration of physiological data that we developed. It was also used as a monitoring application to ensure that physiological data was being transmitted correctly to the game.}
    \label{fig:calibrationapp}
  \end{center}
\end{figure}

Towards the end of development we quickly felt the need to
normalize the physiological devices' data feed to the
game – this provided an easy abstraction to work over as
we were not working on each person's unique physiological
values any more, but rather on the expected behaviour of each
of the physiological devices across a general population.
For this purpose, we created a very simple calibration
application in \emph{WinForms} in which we established the
maximum and minimum values for each physiological
sensor (Figure~\ref{fig:calibrationapp}). With the exception of the TEMP sensor,
all sensors were calibrated for each person before a
playtesting session.  EMG-Legs, EMG-Arm and RESP were
normalized to the [0.0, 1.0] range. The TEMP signal passed to the game only has two states: on and off, represented by a byte of value 0 or 255, respectively. Unlike the other sensors, this one is
dynamically post-processed inside the \emph{BiofeedbackDLL} to achieve the toggle-like
behaviour.

The raw TEMP data
indicates the actual temperature of whatever object the sensor is touching: since
in our interaction model it is attached to a headset (in the place of a microphone),
the sensor registers room temperature values. To detect when the player is blowing
hot air on the sensor (i.e. a temperature difference), we compute an approximation
of the TEMP's first derivative by using the current and previous readings of the
raw TEMP sensor, divided by the time difference between different game frames (``delta time'' in the game).
When this instant derivative crosses certain threshold values, determined by trial-and-error,
we change the button state of the TEMP sensor.

\begin{figure}
  \begin{center}
    \leavevmode
    \includegraphics[width=0.50\textwidth]{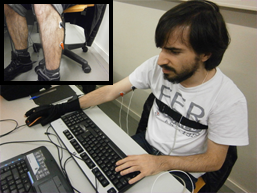}
    \caption[Beta-tester playing our game with physiological devices]{Example of a beta-tester of \emph{``GenericShooter 3000''} with the physiological devices equipped. In this image the following sensors are visible: EMG-Arm, 2x EMG-Leg (top-left corner), RESP and GLOVE. The TEMP sensor was attached to the tip of a headset equipped by the players.}
    \label{fig:subject}
  \end{center}
\end{figure}

For the GLOVE device we used automatic calibration and
gesture detection capabilities existing in the
manufacturer's library – a total of 15 hand poses can
be detected by the device. Figure~\ref{fig:subject} shows how the
equipment was connected to players.

\section{Iterative Playtesting and Game Tutorial}

To assure the game was played as our game designer intended
and had the intended quality in gameplay comparison with
existing FPS titles, 8 voluntaries were called for individual
pilot playtesting sessions over the course of development as
suggested in~\cite{fullerton2008game}. This was useful to eliminate
small bugs, but also to fine-tune gameplay details and level
design issues in all three versions of the game.

\begin{figure}
  \begin{center}
    \leavevmode
    \includegraphics[width=1.00\textwidth]{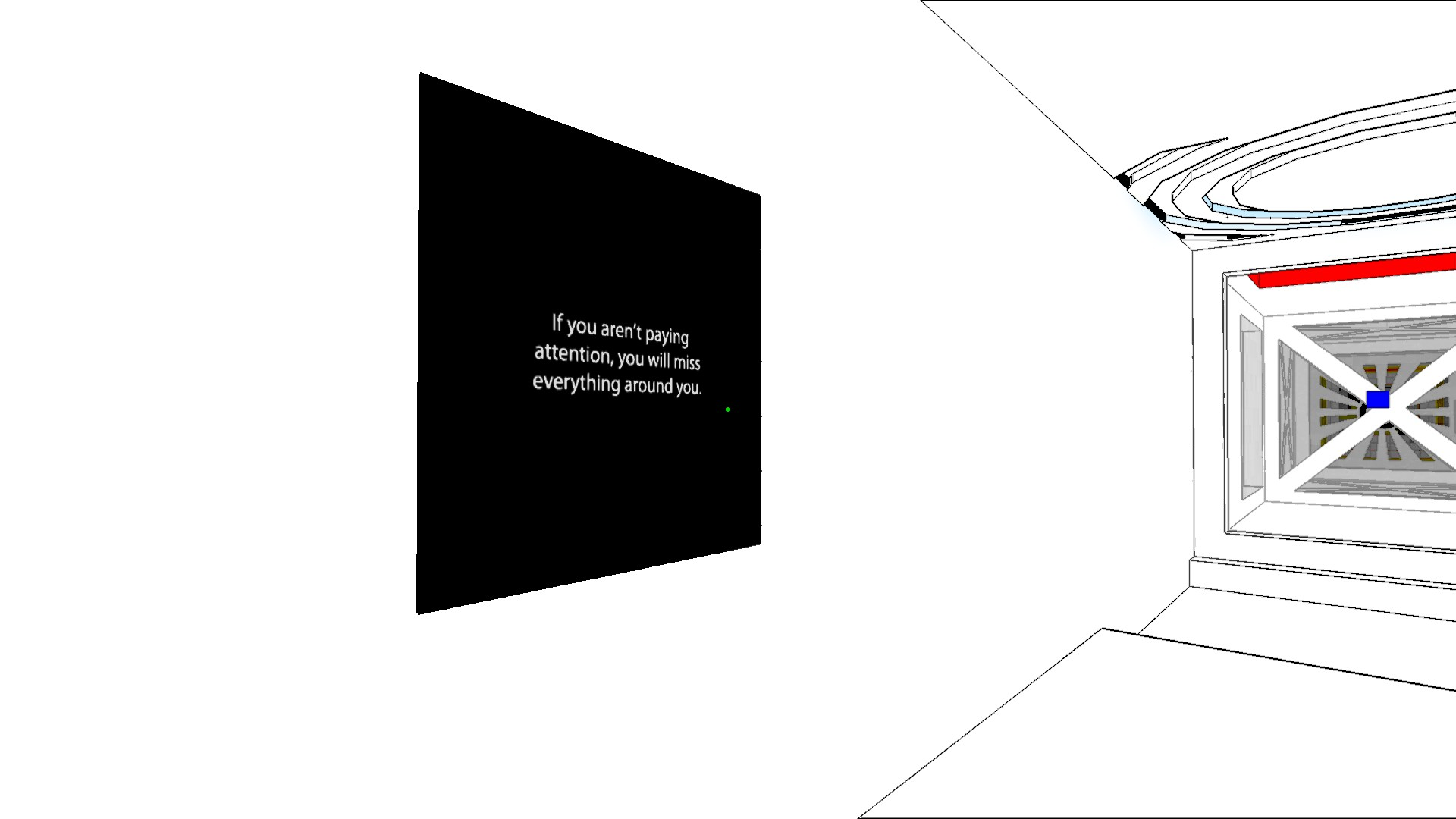}
    \caption[Hint message example from the game ``Antichamber'']{Example of a hint message from the game \emph{``Antichamber''}. The message reads \emph{``If you aren't paying attention, you will miss everything around you.''}, which alerts the player to be on the lookout for hidden rooms or secret passages nearby.}
    \label{fig:antichamber}
  \end{center}
\end{figure}

\begin{figure}
  \begin{center}
    \leavevmode
    \includegraphics[width=1.00\textwidth]{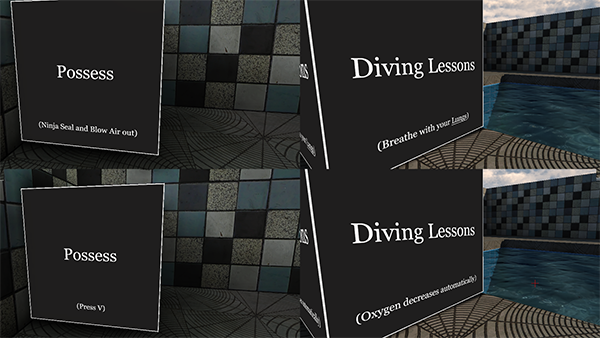}
    \caption[Hint box examples in our Tutorial level]{Examples of hint boxes in the Tutorial level. Vanilla versions (bottom), Multimodal Possess (top-left) and Biofeedback Underwater Breathing (top-right).}
    \label{fig:tutorial}
  \end{center}
\end{figure}

Early testing also showed that despite FPS games being a
popular genre, some of our game testers always died at the
beginning of the prison level. To eliminate this issue – partly
inspired by the strong tutorial present in the game
\emph{``Age of Empires 2: Age of Kings''} and a testimony from a
researcher working in that same game~\cite[chap.\ 9, p. 267]{fullerton2008game} – we
created a 5-minute playable tutorial level for the three
versions where players could fully understand the game
and what their character is able to do. Hint boxes
inspired by the game \emph{``Antichamber''} (Figure~\ref{fig:antichamber}) were placed in key points
of the tutorial level\footnote{Floor and mosaic wall textures created by Hugues Muller - available at http://nobiax.deviantart.com/gallery/\#Packs.} (Figure~\ref{fig:tutorial}) to help the players to activate all
mechanisms on their own. From their improved
results (none struggled further with the game mechanics)
this proved to be a very effective way to introduce both
the game on its entirety and how the biofeedback sensors
worked as well.

After our \emph{``GenericShooter 3000''} game was finished and
was sufficiently refined, we proceeded to conduct our
empirical study. To recall our research questions, we are
interested in accessing:

\begin{enumerate}
\item The players' reactions to physiologically augmented gameplay mechanisms in modern FPS shooters;
\item A comparison between multimodal and unimodal mechanisms in terms of in-game action performance, situation context and player experience;
\item Whether we can enhance game realism and depth using multimodal and unimodal biofeedback.
\end{enumerate}

\chapter{An Empirical Study with Players}\label{chap:chap4}

\section{Game Conditions}

The study was comprised of three conditions (\textbf{Vanilla},
\textbf{Unimodal} and \textbf{Multimodal}) represented by the
three game versions described in the previous section. Only the
latter two used physiological input. Table~\ref{tab:sensorsmechanics}
synthesizes the mapping of sensors for each of the eight
biofeedback-augmented gameplay mechanics.

\begin{table}[t]
  \centering
  \caption{Mapping of Sensors to Game Mechanics}
\begin{tabular}{ lclcl }

       \textbf{Mechanism} & \phantom{abc} & \textbf{Unimodal} & \phantom{abc} & \textbf{Multimodal}\\
	\toprule
        Gun Recoil    && EMG-Arm && EMG-Arm + RESP     \\
        Invisibility  && RESP    && GLOVE + RESP       \\
        Underwater    && \multicolumn{3}{c}{--- RESP - same in both versions ---} \\
        Possession    && TEMP    && GLOVE + TEMP       \\
        Fire Blow     && TEMP    && RESP + TEMP        \\
        Sprinting     && EMG-Leg && 2x EMG-Leg         \\
        Item Use      && GLOVE   && EMG-Arm + GLOVE    \\
        Grab Object   && GLOVE   && EMG-Arm + GLOVE    \\
	\bottomrule
\end{tabular}
  \label{tab:sensorsmechanics}
\end{table}

\section{Experimental Protocol} \label{sec:protocol}

For the study, a three-condition repeated-measures within-subjects
design was used. To mitigate order bias effects, each participant
played the three conditions on a random order. Test sessions were
split in two consecutive parts: Tutorial and Prison Level. Upon
arrival, participants were acknowledged for their time and asked
to sign a consent form. Afterwards, they were equipped with the
physiological devices with a simultaneous briefing on what each
sensor was meant to do. The physiological devices were calibrated
once before the Tutorial section and recalibrated if necessary at
the start of the Prison Level section.

In the first test round, participants played the Tutorial level
three times (one for each test condition). By doing so they were
able to become accustomed to the game and also to experiment
with the different mechanisms from Table~\ref{tab:sensorsmechanics} as they saw fit.
At the end of the third Tutorial, players filled a questionnaire
where they were asked to compare all mechanisms between the three
conditions according to ratings of Fun, Ease of Use and Originality.
This allowed us to get valuable insights on which mechanisms had a
higher potential to add depth to the game experience (Fun), which
were causing issues or improving gameplay (Ease of Use) and which
had a highest impact in the overall gameplay. It also became an
important step in evaluating our game design choices in the
mechanisms themselves.

In a similar way to the word list prompt presented in~\cite{gow2010capturing}, we also asked players to compare each
mechanism between the three conditions using the list of keywords
displayed on Table~\ref{tab:keywordprompt}. However, we designed
ours with a moderate
number of keywords to broadly and swiftly capture differences in
positive and negative aspects of gameplay – allowing us to gain
a significant insight on their game experience with and without
the physiological devices.

\begin{table}[t]
  \centering
  \newcommand{\cc}{\centering\arraybackslash}
  \newcommand{\tn}{\tabularnewline}
  \caption{Card Sorting - Describing Game Mechanics in Words}
  \begin{tabular}{>{\cc}m{2.5cm} l >{\cc}m{2.5cm} l >{\cc}m{2.5cm} l >{\cc}m{2.5cm} @{}l}
	\toprule[0.12em]
      Useless in-game && Simple && Realistic && Exhausting &  \\[0.5cm]
      Complete (``don't mess with it'') && Imaginary && Core in-game && Complex  & \tn[0.5cm]
      Confusing && Relaxing && Intuitive && Incomplete (``something's missing'') & \tn[0.5cm]
	\bottomrule[0.12em]
  \end{tabular}
  \label{tab:keywordprompt}
\end{table}

In the second test round, participants played the Prison Level
three times for each condition until they managed to escape
the prison (approximately 15-20 minutes in the initial playthroughs and
5-15 minutes on subsequent runs). At the end of the third
Prison Level, players compared each version of the game
(Vanilla vs Unimodal vs Multimodal) according to ratings of
Originality, Preference and Playability.
Playability was inspired in Brown's immersion theory where game controls
were an element of the first barrier to immersion, ``Accessibility''~\cite{brown2004grounded}.
Being a rather ambiguous term, in the context of our study we presented it to our participants
as \emph{``the degree to which they felt the game controls were an
obstacle in each condition''} (in simpler terms, the quality of game
controls). They were also required to comment their rating choices
to justify their opinion and avoid randomly-filled questions.

Lastly, they filled an Intrinsic Motivation Inventory (IMI)
Post Experience Questionnaire with each question repeated
three times to compare the game versions. Players were
ultimately acknowledged again for their participation and
received a chocolate bar as a reward.

\section{Testing Apparatus}

The game was played on a 64-bit desktop computer at a 1680x1050
resolution, running Windows 7 Enterprise SP 1 with the following
hardware specifications:

\begin{itemize}
\item Intel® Core™2 Quad Q9550 @ 2.83 GHz
\item 4 GB RAM
\item NVIDIA® GeForce® 9800 GTX
\item Monitor ASUS VW222S with a size of 22 inches
\item 2 desktop speakers of small size
\end{itemize}

Physiological data was captured using the NeXus-10 device
and the 5DT Glove, and integrated in the game using our
custom DLL as described in the previous section. The
machine's frame rates were logged for all three
conditions to monitor performance issues that might
detract player experience. Luckily, no significant
(>5 frames per second) frame rate differences were
observed.

\section{Participants' Demographic Data}

\begin{figure}[t]
  \begin{center}
    \leavevmode
    \includegraphics[width=1.00\textwidth]{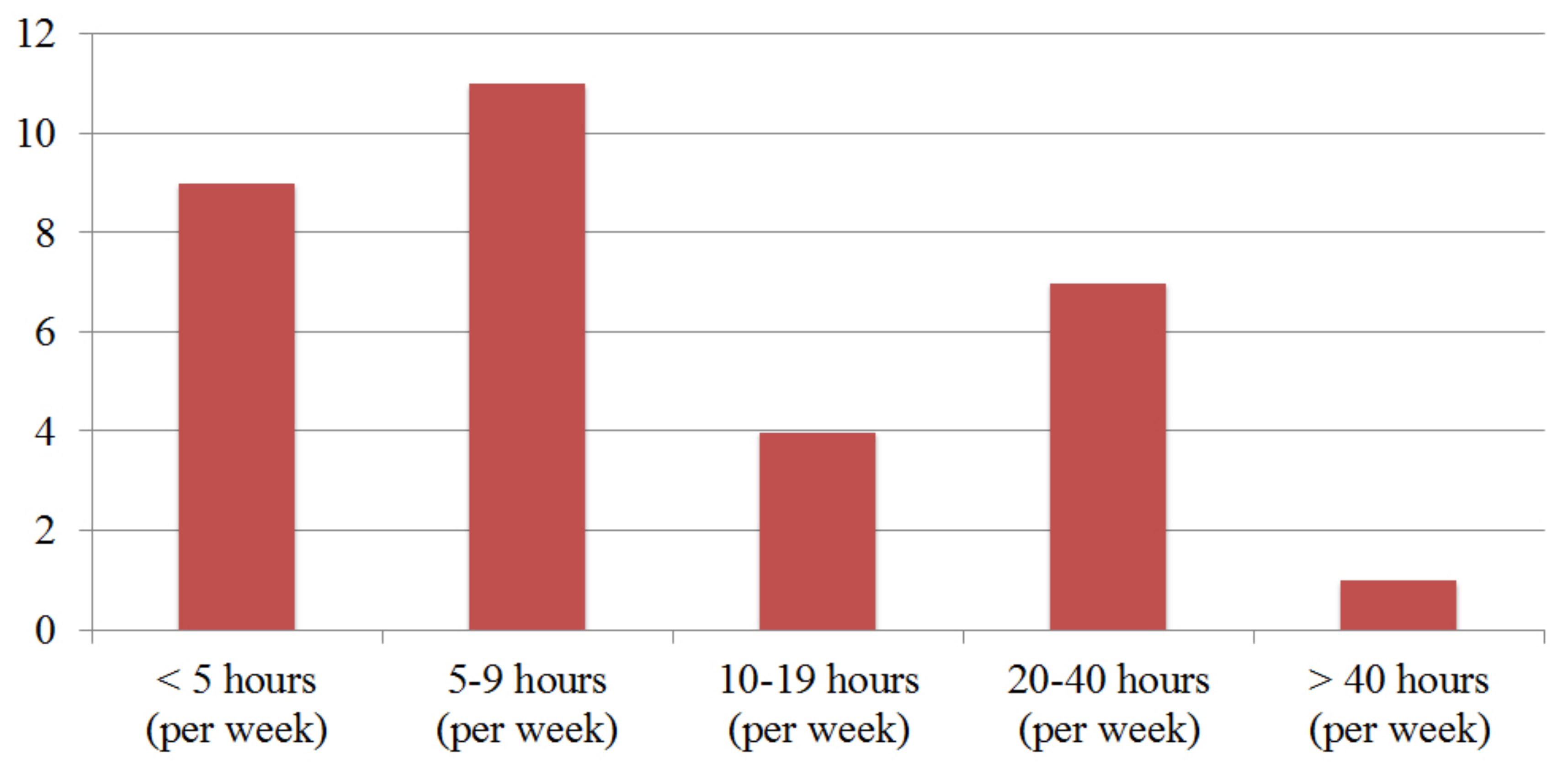}
    \caption[Participants' average time spent playing videogames]{The participants' average time spent playing videogames.}
    \label{fig:hours}
  \end{center}
\end{figure}

Thirty-two participants (29 male, 3 female) aged from
18 to 27 (M=21.28, SD=2.56) participated in the study – before
participating in the study sessions they were required to
fill a demographics questionnaire. Figure~\ref{fig:hours} depicts the
average number of hours per week our participants spent
playing videogames.

\begin{figure}[t]
  \begin{center}
    \leavevmode
    \includegraphics[width=1.00\textwidth]{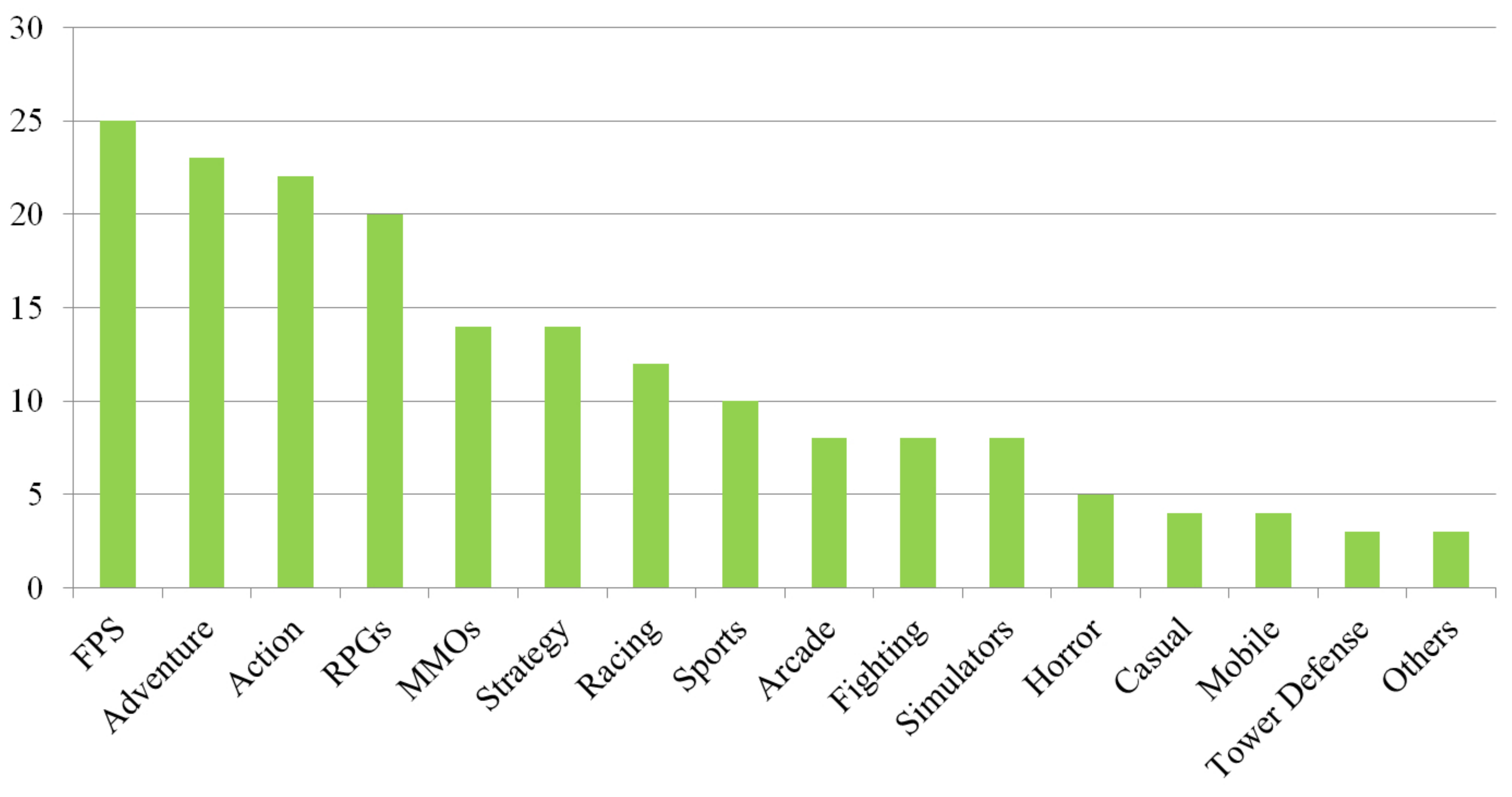}
    \caption[Game genres usually played by our participants]{Game genres usually played by our participants.}
    \label{fig:genres}
  \end{center}
\end{figure}

Players have also reported which game genres they usually
play (Figure~\ref{fig:genres}) - FPS, Adventure and Action genres were consistently
mentioned among players, so we can assume our players to be
fairly familiar (even if no proficient) with the genre.

\chapter{Empiric Study Results} \label{chap:chapter5}

\section*{}

This section is divided in two main sub-sections. The first of
these analyses how participants rated each of the individual
gameplay mechanics and how they differed between their variants.
Ratings regarding the Fun, Ease of Use and Originality aspects
were processed using One-way Analysis of Variance (ANOVA) tests
with the Vanilla, Unimodal and Multimodal versions of the game
as the within-subjects factor. Post-hoc Tukey tests were performed
when statistical significance was met.

For the list of evaluation keywords describing the gameplay
mechanics, we present a comparison of the most frequent keywords
between the different mechanisms. This is meant as a more
subjective and preliminary insight on our participants' game
experience.

In sub-section~\ref{sec:quizconditions} we focus our analysis on how participants
perceived the overall gameplay experience to be affected across
the three gameplay conditions in respect to the components of
Originality, Preference and Playability. Finally, for a more
objective analysis of player experience provided by each of the
game conditions, we resorted to the Intrinsic Motivation Inventory (IMI)
questionnaire. Here, the measured components were:
Interest/Enjoyment, Perceived Competence, Effort/Importance,
Pressure/Tension, Perceived Choice, Value/Usefulness and Relatedness.

\section{Questionnaire 1: Individual Mechanics} \label{sec:quizmechanics}
\subsection{Fun Ratings}

\begin{figure}[t]
  \begin{center}
    \leavevmode
    \includegraphics[width=1.00\textwidth]{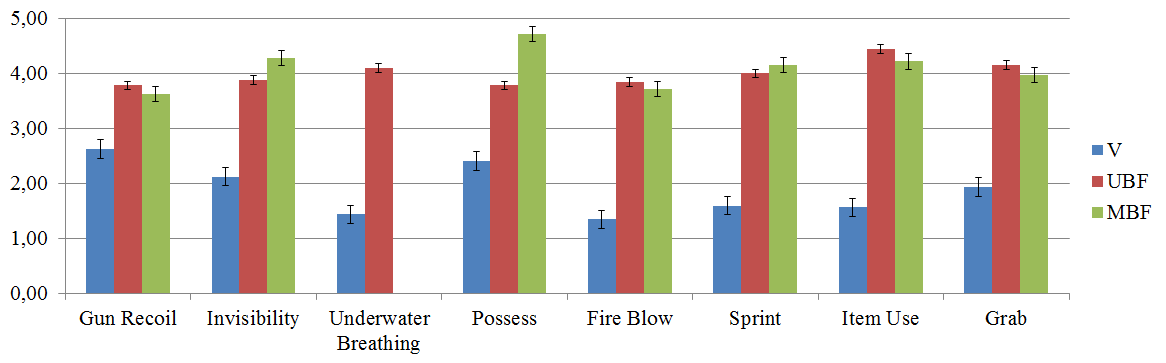}
    \caption[Q1 Biofeedback Mechanics - Mean Fun with Standard Error]{Mean Fun values between conditions with corresponding Standard Error measures.}
    \label{fig:q1fun}
  \end{center}
\end{figure}

In the first questionnaire we asked players to rate the Fun
of each mechanism (Figure~\ref{fig:q1fun}) on a scale of 0 (Not fun) to 5(Very fun).
For the mechanisms Gun Recoil ($\chi^2(2) = 0.856, p > 0.01$);
Invisibility ($\chi^2(2) = 0.805, p > 0.01$);
Underwater (it only has two versions, by definition it does not violate
the sphericity assumption); Fire Blow ($\chi^2(2) = 0.950, p > 0.01$);
Item Use ($\chi^2(2) = 0.996, p > 0.01$) and Grab ($\chi^2(2) = 0.756, p > 0.01$),
Mauchly’s sphericity assumption had been met. For Possess ($\chi^2(2) = 0.883, p < 0.01$)
and Sprint ($\chi^2(2) = 0.644, p < 0.01$) it was violated. As a result,
degrees of freedom for both mechanisms were corrected using
Greenhouse-Geisser estimates of sphericity
($\varepsilon = 0.71562$ and $\varepsilon = 0.73746$, respectively).

\begin{table}[t]
  \centering
  \caption{Q1 Fun - Statistical Analysis}
\begin{tabular}{ l  rr c rrr }
    & \multicolumn{2}{c}{One-Way ANOVA} & \phantom{abc} & \multicolumn{3}{c}{Tukey Post-Hoc} \\
    \cmidrule{2-3} \cmidrule{5-7}
    \textbf{\textit{Fun}} & $F$ & $p$ && V-UBF & V-MBF & UBF-MBF \\
	\toprule[0.15em]
       Gun Recoil           &  13.584 & < 0.001 && < 0.001 & < 0.001 & 0.794 \\
       Invisibility         &  61.076 & < 0.001 && < 0.001 & < 0.001 & 0.131 \\
       Underwater Breathing & 111.330 & < 0.001 && \multicolumn{3}{c}{---} \\
       Possession           &  64.807 & < 0.001 && < 0.001 & < 0.001 & < 0.001 \\
       Fire Blow            & 77.196  & < 0.001 && < 0.001 & < 0.001 & 0.846 \\
       Sprint               & 74.975  & < 0.001 && < 0.001 & < 0.001 & 0.784 \\
       Item Use             & 96.380  & < 0.001 && < 0.001 & < 0.001 & 0.612 \\
       Grab                 & 56.564  & < 0.001 && < 0.001 & < 0.001 & 0.698 \\
	\bottomrule[0.15em]
\end{tabular}
  \label{tab:q1funanalysis}
\end{table}

Statistical significance was achieved for all mechanisms:
Gun Recoil $F(2,62) = 13.584, p < 0.01$; Invisibility $F(2,62) = 61.076, p < 0.01$;
Underwater $F(1,31) = 114.33, p < 0.01$; Possession $F(2,62) = 64.807, p < 0.01$;
Fire Blow $F(2,62) = 77.196, p < 0.01$; Sprint $F(2,62) = 74.975, p < 0.01$;
Item Use $F(2,62) = 96.38, p < 0.01$; Grab $F(2,62) = 56.564, p < 0.01$.

Tukey post-huc tests revealed that participants found both the
unimodal and multimodal biofeedback versions of the eight mechanics
were more fun than the vanilla version of the game ($p < 0.01$).
Possession's multimodal version was the only mechanism that players
considered more fun than the unimodal version ($p < 0.01$). No further
differences were detected in the remaining mechanisms ($p > 0.01$).

The Underwater mechanism did not require a Tukey test nor a
Mauchly's sphericity test as it only existed in two versions:
vanilla and biofeedback control. It was considered more Fun in the
biofeedback control version than the vanilla version.

\subsection{Ease of Use Ratings}

\begin{figure}[t]
  \begin{center}
    \leavevmode
    \includegraphics[width=1.00\textwidth]{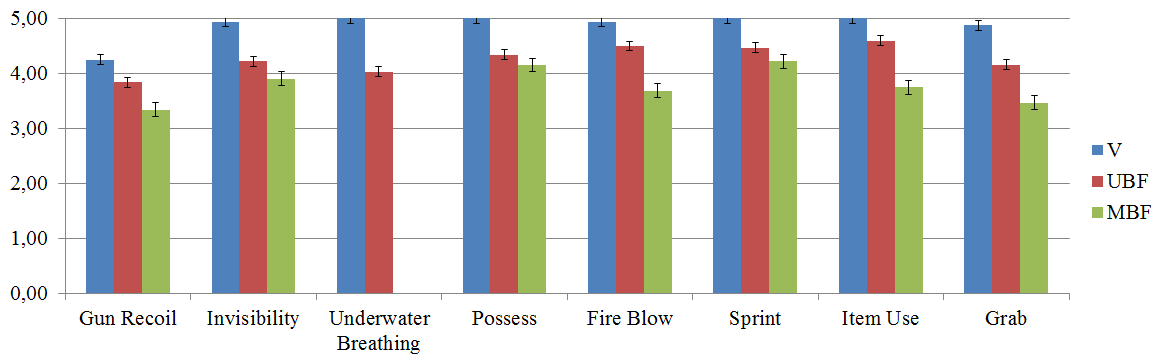}
    \caption[Q1 Biofeedback Mechanics - Mean Ease of Use between conditions with Standard Error]{Mean Ease of Use values between conditions with corresponding Standard Error measures.}
    \label{fig:q1ease}
  \end{center}
\end{figure}

Participants were asked to rate each mechanism also regarding their
Ease of Use (Figure~\ref{fig:q1ease}) from 0 (Very hard) to 5 (Very easy). For all
eight mechanisms Mauchly’s sphericity assumption was met:
Gun Recoil ($\chi^2(2) = 0.955, p > 0.01$); Invisibility ($\chi^2(2) = 0.870, p > 0.01$);
Underwater; Possess ($\chi 2(2) = 0.883, p > 0.01$); Fire Blow ($\chi^2(2) = 0.830, p > 0.01$);
Sprint ($\chi^2(2) = 0.994, p > 0.01$); Item Use ($\chi^2(2) = 0.814, p > 0.01$)
and Grab ($\chi^2(2) = 0.874, p > 0.01$).

\begin{table}[t]
  \centering
  \caption{Q1 Ease of Use - Statistical Analysis}
\begin{tabular}{ l  rr c rrr }
    & \multicolumn{2}{c}{One-Way ANOVA} & \phantom{abc} & \multicolumn{3}{c}{Tukey Post-Hoc} \\
    \cmidrule{2-3} \cmidrule{5-7}
    \textbf{\textit{Ease of Use}} & $F$ & $p$ && V-UBF & V-MBF & UBF-MBF \\
	\toprule[0.15em]
       Gun Recoil           &  7.281  & < 0.01 && 0.210 & < 0.001 & 0.098 \\
       Invisibility         & 24.363  & < 0.001 && < 0.001 & < 0.001 & 0.106 \\
       Underwater Breathing & 32.137  & < 0.001 && \multicolumn{3}{c}{---} \\
       Possession           & 28.981  & < 0.001 && < 0.001 & < 0.001 & 0.249 \\
       Fire Blow            & 28.257  & < 0.001 && 0.031 & < 0.001 & < 0.001 \\
       Sprint               & 20.851  & < 0.001 && < 0.001 & < 0.001 & 0.115 \\
       Item Use             & 40.374  & < 0.001 && 0.156 & < 0.001 & < 0.001 \\
       Grab                 & 46.669  & < 0.001 && < 0.001 & < 0.001 & < 0.001 \\
	\bottomrule[0.15em]
\end{tabular}
  \label{tab:q1easeanalysis}
\end{table}

Statistical significance was achieved as well for all eight mechanisms:
Gun Recoil $F(2,62) = 7.281, p < 0.01$; Invisibility $F(2,62) = 24.363, p < 0.01$;
Underwater $F(1,31) = 32.137, p < 0.01$; Possess $F(2,62) = 28.981, p < 0.01$;
Fire Blow $F(2,62) = 28.257, p < 0.01$; Sprint $F(2,62) = 20.851, p < 0.01$;
Item Use $F(2,62) = 40.374, p < 0.01$; Grab $F(2,62) = 46.669, p < 0.01$.

Tukey post-hoc tests showed that players:

\begin{enumerate}
\item Found the vanilla versions easier to use than the unimodal
versions for Invisibility, Possess, Sprint and Grab ($p < 0.01$);
\item Found all mechanisms easier to use in the vanilla version
than the multimodal version ($p < 0.01$)
\item Found Fire Blow, Use and Grab easier to use in the unimodal
version than the multimodal version ($p < 0.01$).
\end{enumerate}

No differences were detected for the remaining combinations ($p > 0.01$).

The Underwater Breathing mechanic did not require a Tukey test nor
a Mauchly's sphericity test as it only existed in two versions:
vanilla and biofeedback control. It was considered easier to use
in the vanilla version than the biofeedback version.

\subsection{Originality Ratings}

\begin{figure}[t]
  \begin{center}
    \leavevmode
    \includegraphics[width=1.00\textwidth]{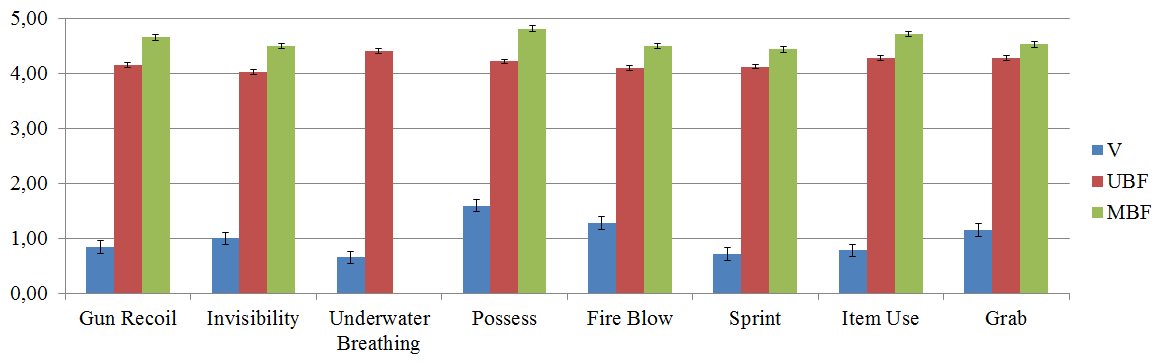}
    \caption[Q1 Biofeedback Mechanics - Mean Originality between conditions with Standard Error]{Mean Originality values between conditions with corresponding Standard Error measures.}
    \label{fig:q1originality}
  \end{center}
\end{figure}

Originality was rated on a scale of 0 (No originality) to 5 (Very original)
– the results are summarized on Figure~\ref{fig:q1originality}. For the mechanisms
Invisibility ($\chi^2(2) = 0.746, p > 0.01$) and Underwater Mauchly's
sphericity assumption was met. For Gun Recoil ($\chi^2(2) = 0.724, p < 0.01$);
Possess ($\chi^2(2) = 0.471, p < 0.01$); Fire Blow ($\chi^2(2) = 0.414, p < 0.01$);
Sprint ($\chi^2(2) = 0.473, p < 0.01$); Item Use ($\chi^2(2) = 0.636, p < 0.01$) and
Grab ($\chi^2(2) = 0.504, p < 0.01$) it was violated. Greenhouse-Geisser
corrections were applied in those cases
($\varepsilon = 0.78344$; $0.65385$; $0.63047$; $0.65496$; $0.73294$ and $0.66852$, respectively).

\begin{table}[t]
  \centering
  \caption{Q1 Originality - Statistical Analysis}
\begin{tabular}{ l  rr c rrr }
    & \multicolumn{2}{c}{One-Way ANOVA} & \phantom{abc} & \multicolumn{3}{c}{Tukey Post-Hoc} \\
    \cmidrule{2-3} \cmidrule{5-7}
    \textbf{\textit{Originality}} & $F$ & $p$ && V-UBF & V-MBF & UBF-MBF \\
	\toprule[0.15em]
       Gun Recoil           & 203.19  & < 0.001 && < 0.001 & < 0.001 & 0.046 \\
       Invisibility         & 183.73  & < 0.001 && < 0.001 & < 0.001 & 0.054 \\
       Underwater Breathing & 317.05  & < 0.001 && \multicolumn{3}{c}{---} \\
       Possession           & 127.76  & < 0.001 && < 0.001 & < 0.001 & 0.020 \\
       Fire Blow            & 154.90  & < 0.001 && < 0.001 & < 0.001 & 0.111 \\
       Sprint               & 236.54  & < 0.001 && < 0.001 & < 0.001 & 0.234 \\
       Item Use             & 339.10  & < 0.001 && < 0.001 & < 0.001 & 0.028 \\
       Grab                 & 160.68  & < 0.001 && < 0.001 & < 0.001 & 0.463 \\
	\bottomrule[0.15em]
\end{tabular}
  \label{tab:q1originanalysis}
\end{table}

Statistical significance was achieved for all eight mechanisms:
Gun Recoil $F(2,62) = 203.19, p < 0.01$; Invisibility $F(2,62) = 183.73, p < 0.01$;
Underwater $F(1,31) = 317.05, p < 0.01$; Possess $F(2,62) = 127.76, p < 0.01$;
Fire Blow $F(2,62) = 154.90, p < 0.01$; Sprint $F(2,62) = 236.54, p < 0.01$;
Item Use $F(2,62) = 339.10, p < 0.01$; Grab $F(2,62) = 160.68, p < 0.01$.

Tukey post-hoc tests revealed that players found both biofeedback
versions more original than the vanilla version ($p < 0.01$).
No differences were detected between the unimodal and multimodal
versions ($p > 0.01$).

The Underwater Breathing mechanic did not require a Tukey test nor a
Mauchly's sphericity test as it only existed in two versions:
vanilla and biofeedback control. Players found that the Underwater
biofeedback mechanic was clearly more original than its
vanilla counterpart.

\section{Gameplay Keywords}

Tables~\ref{tab:keywordresult1}-~\ref{tab:keywordresult2}
show the results of the participants' evaluation of each game mechanic (in each version) using the presented list of keywords. We will be discussing only relevant results of the table as the amount of information is too
extensive for a complete discussion. However, the results are presented as a whole so they can be used as a reference and comparison basis for future work on new direct biofeedback games and game design experiments.
To view the gathered data in a more friendly format (column charts), please refer to Appendix~\ref{ap1:charts}.

\begin{sidewaystable}[htbf]
  \centering
  \caption{Card Sorting Results - Describing Game Mechanics in Words (1)}
\begin{tabular}{l rrr r rrr r rrr r rrr}
	% \toprule
       & \multicolumn{3}{c}{Sprint} & \phantom{abc}
       & \multicolumn{3}{c}{Gun Recoil} & \phantom{abc}
       & \multicolumn{2}{c}{ \begin{tabular}{@{}c@{}}Underwater\\Breathing\end{tabular} } & \phantom{abc}
       % & \multicolumn{2}{c}{\shortstack[c]{Underwater \\ Breathing}} & \phantom{abc}
       & \multicolumn{3}{c}{Invisibility} \\
       \cmidrule{2-4} \cmidrule{6-8} \cmidrule{10-11} \cmidrule{13-15}
       & Vanilla & UBF & MBF && Vanilla & UBF & MBF && Vanilla & BF && Vanilla & UBF & MBF \\
    \midrule[0.15em]

       \textbf{Useless in-game} & 0\% & 3\% & 3\% && 0\% & 0\% & 3\% && 13\% & 0\% && 3\% & 3\% & 9\% \\
       \textbf{Core in-game} & 28\% & 34\% & 44\% && 19\% & 16\% & 16\% && 9\% & 19\% && 19\% & 34\% & 31\% \\
     \midrule
       \textbf{Simple}  & 100\% & 78\% & 56\% && 84\% & 50\% &  6\% && 84\% & 50\% && 97\% & 47\% & 31\% \\
       \textbf{Complex} &   0\% &  0\% &  3\% &&  3\% & 16\% & 41\% &&  0\% &  0\% && 0\%  &  9\%  & 34\% \\
     \midrule
       \textbf{Realistic} & 6\% & 31\% & 63\% && 3\% & 69\% & 72\% && 9\% & 84\% &&  0\% & 38\% & 38\% \\
       \textbf{Imaginary} & 9\% &  9\% &  9\% && 0\% &  0\% & 0\%  && 6\% &  6\% && 13\% & 34\% & 38\% \\
     \midrule
       \textbf{Exhaustive} & 0\% & 9\% & 28\% && 6\% & 28\% & 41\% && 0\% & 22\% && 0\% & 16\% & 19\% \\
       \textbf{Relaxed} & 41\% & 25\% & 16\% && 31\% & 13\% & 3\% && 22\% & 3\% && 31\% & 34\% & 19\% \\
     \midrule
       \textbf{Confusing} &  0\% &  9\% &  3\% &&  0\% &  9\% & 28\% &&  0\% &  3\% &&  0\% &  3\% & 6\% \\
       \textbf{Intuitive} & 38\% & 50\% & 63\% && 53\% & 53\% & 44\% && 22\% & 97\% && 28\% & 41\% & 44\% \\
     \midrule
       \textbf{Incomplete} & 22\% & 16\% & 9\% && 22\% & 34\% & 16\% && 25\% & 3\% && 16\% & 31\% & 3\% \\
       \textbf{Complete} & 19\% & 22\% & 31\% && 22\% & 9\% & 28\% && 16\% & 66\% && 22\% & 25\% & 44\% \\
	\bottomrule[0.15em]
\end{tabular}
  \label{tab:keywordresult1}
\end{sidewaystable}

\begin{sidewaystable}[htbf]
  \centering
  \caption{Card Sorting Results - Describing Game Mechanics in Words (2)}
\begin{tabular}{l rrr r rrr r rrr r rrr}
       & \multicolumn{3}{c}{Item Use} & \phantom{abc}
       & \multicolumn{3}{c}{Possession} & \phantom{abc}
       & \multicolumn{3}{c}{Grab} & \phantom{abc}
       & \multicolumn{3}{c}{Fire Blow} \\
       \cmidrule{2-4} \cmidrule{6-8} \cmidrule{10-12} \cmidrule{14-16}
       & Vanilla & UBF & MBF && Vanilla & UBF & MBF && Vanilla & UBF & MBF && Vanilla & UBF & MBF \\
  \midrule[0.15em]

    \textbf{Useless in-game} & 0\% & 0\% & 3\% && 0\% & 0\% & 0\% && 0\% & 0\% & 0\% && 19\% & 13\% & 13\% \\
    \textbf{Core in-game} & 25\% & 31\% & 31\%&& 25\% & 28\% & 38\% && 25\% & 34\% & 34\% && 13\% & 9\% & 9\% \\
  \midrule
    \textbf{Simple} &100\% & 75\% & 44\% && 94\% & 69\% & 34\% && 94\% & 56\% & 31\% && 88\% & 78\% & 34\% \\
    \textbf{Complex}&  0\% &  0\% & 34\% &&  0\% &  6\% & 25\% && 3\% &  6\% & 19\% && 0\%  & 0\% & 28\% \\
  \midrule
    \textbf{Realistic} & 3\% & 66\% & 81\% && 0\% & 13\% & 22\% && 3\% & 66\% & 69\% && 13\% & 78\% & 78\% \\
    \textbf{Imaginary} & 3\% &  3\% &  6\% &&31\% & 53\% & 53\% &&16\% &  6\% & 6\% && 6\% & 3\% & 3\% \\
  \midrule
    \textbf{Exhaustive} & 0\% & 3\% & 22\% && 0\% & 3\% & 6\% && 0\% & 6\% & 16\% && 0\% & 0\% & 31\% \\
    \textbf{Relaxed}& 31\% & 41\% & 16\% && 19\% & 31\%& 25\% && 25\% & 34\% & 0\% && 31\% & 34\% & 19\% \\
  \midrule
    \textbf{Confusing} & 0\% & 3\% & 13\% && 0\% & 3\% & 6\% && 0\% & 6\% & 25\% && 3\% & 0\% & 0\% \\
    \textbf{Intuitive} &41\% &72\% & 50\% &&31\% & 34\% & 41\%&& 34\% &59\% &59\%&& 25\% & 91\% & 72\% \\
  \midrule
    \textbf{Incomplete} & 31\% & 16\% & 3\% && 38\% & 16\% &3\% && 19\% &28\% &38\% && 22\% & 13\% & 3\% \\
    \textbf{Complete} & 16\% & 31\% & 47\% && 13\% &22\% & 50\% && 19\% & 22\% & 34\% && 16\%& 31\% & 34\% \\
\bottomrule[0.15em]
\end{tabular}
  \label{tab:keywordresult2}
\end{sidewaystable}

The Fire Blow mechanic registered moderate ``Useless'' values
most likely because it was only used as a puzzle-solving ability
and did not have much influence in the main action sections of
the game. However, we are certain that its evolution to a
stealth-related mechanic in the game (see Section~\ref{ssec:fireblow})
would drop this value to 0\% as it would be more relevant
to the game.

The ``Core in-game'' keyword registered an increasing trend in
the biofeedback versions, although not on a very significant
scale – surprisingly, the multimodal version of Sprint received
this adjective 44\% of the times (against 28\% in the vanilla
version). This provides a hint that this mechanic feels more
like a core aspect of gameplay on the multimodal version of the game.

For the ``Simple/Complex'' pair – and within our expectations – the
biofeedback conditions suffered relevant decreases towards the
vanilla version (especially in the multimodal version) and
increases in complexity towards the multimodal version.
However, ``Complex'' did not always translate to something
necessarily bad according to players' feedback:

\begin{itemize}
\item  \emph{``The biofeedback sections don't make the game hard at all,
  but they make it less fluid compared to the vanilla condition.
  While on the latter you can walk and open doors quite fast, on
  the biofeedback versions you have to stop and use the glove''} (P20) 

\item  \emph{``it's complex in a good way, it has depth''}
   (P15 on multimodal's Gun Recoil) 

\item  \emph{``I liked the unimodal version because we have to use more
  elements of our body to get immersed in the game itself and it
  makes it more interesting. The multimodal version is one step
  further compared to the unimodal version because it brings a
  little more complexity (depth) to the gestures execution''} (P25)

\item  \emph{``It's positive because it gives more depth to the game
  (and more immersion), but other people might prefer something more
  simple''} (P4 on multimodal's Gun Recoil)
\end{itemize}

In terms of Realism, both biofeedback versions had the expected results
and registered large increases towards the vanilla version. During
development phase, underwater breathing was one of the most challenging
actions in the biofeedback-controlled versions and we were very curious
about the players' reaction to it in the final version:

\begin{itemize}
\item  \emph{``I wish the diving section was bigger, it was very short.''} (P4)

\item  \emph{``Sometimes I think it would be nice to feel in a movie or
  game how would it be without breathing''} (P15)

\item  \emph{``The diving system is very creative and the hand seals
  system [of the multimodal version] is as well''} (P2)

\item  \emph{``On the breathing section it was very exhausting – if it
  were a game with very strong underwater components or we had to
  repeat it several times it would be very exhausting.''} (P16)

\item  \emph{``I liked all mechanisms in general except for the
  underwater breathing one.''} (P30)
\end{itemize}

Regarding the ``Imaginary'' keyword, it was a sibling concept for
Realism where a game mechanic could fail to achieve realistic
action (because it does not exist in real-life) but was able
to reproduce quite well how players would go about performing
actions that they could imagine themselves doing (e.g. casting
a spell or becoming invisible). This explains why the special
abilities Invisibility and Possess obtained the highest results
in the ``Imaginary'' parameter – players seem to identify more
with those actions as the they try to tap in the players'
imagination.

Exhaustiveness was also expected to increase in the biofeedback
conditions, although it seems some participants were able to cope
with it during gameplay. It remains to be seen in future studies
with longer testing periods whether the type of biofeedback used
(unimodal or multimodal) or even biofeedback itself have an
important influence on the players' willingness to play.

\section{Questionnaire 2: Game Conditions} \label{sec:quizconditions}

\begin{figure}[t]
  \begin{center}
    \leavevmode
    \includegraphics[width=1.00\textwidth]{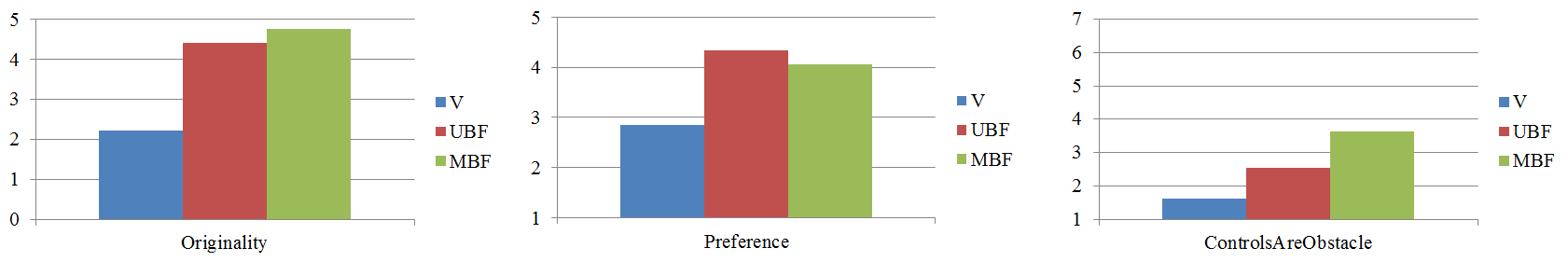}
    \caption[Q2 Game Conditions - Mean Originality, Preference and Playability]{Mean Originality, Preference and Playability values between conditions.}
    \label{fig:q2params}
  \end{center}
\end{figure}

In the second questionnaire participants had to compare the
three game conditions on their entirety instead of evaluating
game mechanics individually. The questionnaire was filled after
all three Prison Levels were completed to limit subjectivity - that is,
by forcing players to compare all three gaming conditions
at the same time, it forced them to reflect on each of them and
weigh them against one another. The comparison of mean scores
for each measured aspect is shown on Figures~\ref{fig:q2params}
and~\ref{fig:q2imi}.

\subsection{Originality, Preference, Playability}

Originality was rated on a scale of 0 (No originality) to
5 (Very original) and Preference on a scale of 1 (Very low) to
5 (Very high). Players were also required to rate the sentence
\emph{``I feel that the game controls were an obstacle in my way to
play well...''}~\cite{brown2004grounded} for each game condition using a seven-level Likert
scale in order to quantify Playability (1 – completely disagree;
4 – neutral; 7 – completely agree). Mauchly’s test statistic
revealed that the sphericity assumption was met for Preference
($\chi^2(2) = 0.933, p > 0.01$) and Playability ($\chi^2(2) = 0.955, p > 0.01$)
but not for Originality ($\chi^2(2) = 0.359, p < 0.01$).
Therefore, Originality’s degrees of freedom were corrected
using a Greenhouse-Geisser estimate of sphericity
($\varepsilon = 0.60937$).

\begin{table}[t]
  \centering
  \caption{Q2 Originality, Preference and Playability between Conditions - Statistical Analysis}
\begin{tabular}{ l  rr c rrr }
    & \multicolumn{2}{c}{One-Way ANOVA} & \phantom{abc} & \multicolumn{3}{c}{Tukey Post-Hoc} \\
    \cmidrule{2-3} \cmidrule{5-7}
    \textbf{\textit{Conditions}} & $F$ & $p$ && V-UBF & V-MBF & UBF-MBF \\
	\toprule[0.15em]
       Originality        & 107.820  & < 0.001 && < 0.001 & < 0.001 & 0.166 \\
       Preference         &  24.905  & < 0.001 && < 0.001 & < 0.001 & 0.432 \\
       Playability        &  23.188  & < 0.001 && 0.009 & < 0.001 & 0.001 \\
	\bottomrule[0.15em]
\end{tabular}
  \label{tab:q2conditionsanalysis}
\end{table}

Statistical significance was encountered for all three
parameters: Originality $F(2,62) = 107.82, p < 0.01$;
Preference $F(2,62) = 24.905, p < 0.01$;
Playability $F(2,62) = 23.188, p < 0.01$.

Tukey post-hoc tests revealed that:

\begin{enumerate}
\item Players found both biofeedback versions more original than
its vanilla counterpart ($p < 0.01$), with no difference detected
between the unimodal and multimodal versions ($p > 0.01$).
This result corroborates what was obtained in the first
questionnaire regarding the various game mechanics.

\item Players preferred both biofeedback versions when compared
to the vanilla version ($p < 0.01$). No significant difference
was observed between the unimodal and multimodal versions ($p > 0.01$).

\item Players disagreed with the statement that game controls
did not stop them from played well – in other words, it means
that the playability of our game controls reached the desired
level. However, Tukey tests indicated that as controls
increased in complexity (vanilla being the least complex
condition, followed by unimodal and then multimodal),
players agreed less with the statement ($p < 0.01$).
\end{enumerate}

\section{IMI Post Experience Questionnaire} \label{sec:quizimi}

The IMI questionnaire that we used has seven parameters:
Interest/Enjoyment, Perceived Competence, Effort/Importance,
Pressure/Tension, Perceived Choice, Value/Usefulness and Relatedness.
For the parameters: Interest/Enjoyment ($\chi^2(2) = 0.753, p > 0.01$);
Perceived Competence ($\chi^2(2) = 0.958, p > 0.01$);
Pressure/Tension ($\chi^2(2) = 0.737, p > 0.01$) and 
Relatedness ($\chi^2(2) = 0.805, p > 0.01$), Mauchly’s sphericity
assumption was met. For Effort/Importance ($\chi^2(2) = 0.533, p < 0.01$);
Perceived Choice ($\chi^2(2) = 0.300, p < 0.01$)
and Value/Usefulness ($\chi^2(2) = 0.633, p < 0.01$) it was violated.
Therefore, degrees of freedom were corrected using Greenhouse-Geisser
estimates of sphericity ($\varepsilon = 0.68157$, $0.58816$ and $0.73176$, respectively).

\begin{sidewaysfigure}[t]
  \begin{center}
    \leavevmode
    \includegraphics[width=1.00\textwidth]{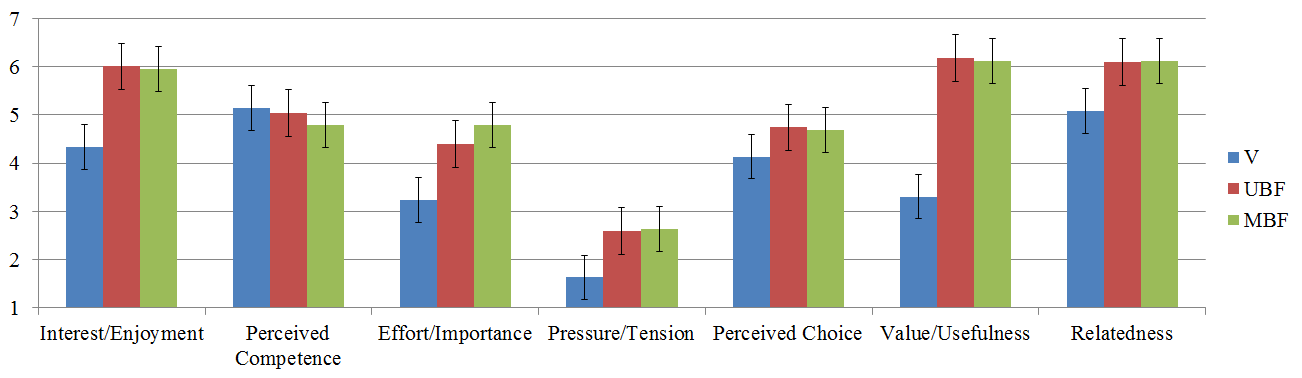}
    \caption[Q2 Game Conditions - Mean IMI parameters with Standard Error]{Mean IMI parameters between conditions with corresponding Standard Error measures.}
    \label{fig:q2imi}
  \end{center}
\end{sidewaysfigure}

%\begin{figure}[t]
%  \begin{center}
%    \leavevmode
%    \includegraphics[width=1.00\textwidth]{q2imifull}
%    \caption[Q2 Game Conditions - Mean IMI parameters with Standard Error]{Mean IMI parameters between conditions with corresponding Standard Error measures.}
%    \label{fig:q2imi}
%  \end{center}
%\end{figure}

\begin{table}[t]
  \centering
  \caption{Q2 IMI Post Experience - Statistical Analysis}
\begin{tabular}{ l  rr c rrr }
    & \multicolumn{2}{c}{One-Way ANOVA} & \phantom{abc} & \multicolumn{3}{c}{Tukey Post-Hoc} \\
    \cmidrule{2-3} \cmidrule{5-7}
    \textbf{\textit{Conditions}} & $F$ & $p$ && V-UBF & V-MBF & UBF-MBF \\
	\toprule[0.15em]
       Interest/Enjoyment     &  58.903  & < 0.001 && < 0.001 & < 0.001 & 0.965 \\
       Perceived Competence   &   3.275  &   0.044 &&   0.762 &   0.041 & 0.185 \\
       Effort/Importance      &  40.743  & < 0.001 && < 0.001 & < 0.001 & 0.083 \\
       Pressure/Tension       &  25.658  & < 0.001 && < 0.001 & < 0.001 & 0.947 \\
       Perceived Choice       &  10.320  & < 0.001 && < 0.001 &   0.001 & 0.951 \\
       Value/Usefulness       &  92.141  & < 0.001 && < 0.001 & < 0.001 & 0.969 \\
       Relatedness            &  20.763  & < 0.001 && < 0.001 & < 0.001 & 0.996 \\
	\bottomrule[0.15em]
\end{tabular}
  \label{tab:q2imianalysis}
\end{table}

Statistical significance was encountered for:
Interest/Enjoyment $F(2,62) = 58.903, p < 0.01$; Effort/Importance $F(2,62) = 40.743, p < 0.01$; Pressure/Tension $F(2,62) = 25.658, p < 0.01$; Perceived Choice $F(2,62) = 10.320, p < 0.01$;
Value/Usefulness $F(2,62) = 92.141, p < 0.01$; and Relatedness $F(2,62) = 20.763, p < 0.01$ – but not for Perceived Competence $F(2,62) = 3.2746, p > 0.01$.

Tukey post-hoc tests revealed that players' Interest/Enjoyment,
Effort/Importance, Pressure/Tension, Perceived Choice, Value/Usefulness
and Relatedness were higher in both biofeedback conditions when
compared to the vanilla version ($p < 0.01$), with no differences
detected between unimodal and multimodal types ($p > 0.01$).

Regarding the Pressure/Tension parameter increase from the vanilla
conditions to the biofeedback versions, we believe it is connected
with the fact that they were experiencing a new form of interaction
and felt the need to succeed in using the physiological devices.
Although Perceived Competence registered a non-significant descending
trend from the vanilla to the unimodal and multimodal conditions,
we believe this trend is due to the players’ large familiarity
with the keyboard/mouse control. Given that this was their first
experience with physiological control, we could expect similar
levels of perceived competence if players’ had the same level of
familiarity across the three conditions.

We consider the increases from the vanilla to the biofeedback
conditions on the Interest/Enjoyment, Effort/Importance, Perceived
Choice and Relatedness parameters are related with Brown's ``Engagement''
type of immersion~\cite{brown2004grounded}. Players seemed to experience
a rewarding sensation by observing their own physical actions reflected
in the game and thus could relate more with the game – although on a
small but statistically significant scale, players had more freedom
on how to activate biofeedback mechanics rather than just pressing a
button, as depicted in the Perceived Choice columns on Figure~\ref{fig:q2imi}.

Players recognized the potential of this technology to enhance
modern videogames using biofeedback technology, as shown in the
Value/Usefulness parameter. Participant 21 had a very important
view on adapting games to people with health issues or limited mobility:

\begin{itemize}
\item  \emph{``I think that these kinds of movements [biofeedback]
  are good for people with breathing issues or other limiting
  restrictions (...) it would be nice to blend the 3 modes for
  handicapped people.” (''} (P21)
\end{itemize}

The IMI Post-Experience Questionnaire did not detect significant
differences between the unimodal and multimodal types of
interaction. However, we believe this is because the
questionnaire was not designed to detect subtle
differences – a limitation we acknowledged early on and
strived to overcome through the remaining questionnaires
and the Card Sorting process. While the two biofeedback
versions do not have many differences in functionality, we
believe that there are relevant differences to them in terms
of the in-game context they should be used on (for a more in-depth
discussion on this, see section~\ref{sec:univsmulti}).

Having presented the obtained results, in the following section
we discuss their meaning as a whole towards the gameplay
experience and further elaborate on the relevant differences
between unimodal and multimodal biofeedback.

\chapter{Results' Analysis and Discussion} \label{chap:chapter6}

\section*{}

We introduced a new variety of Direct Biofeedback in games
that had not been explored before – Multimodal Direct
Biofeedback – and theorize on the possible differences
that it can bring for biofeedback games. We also analyse
empirically our test game to provide a comprehensive case
study on biofeedback games versus the current mouse/keyboard
control schemes.

\section{Fun Levels in Biofeedback Control} \label{sec:funlevels}

The Fun and Interest/Enjoyment metrics show strong
evidence that players enjoyed playing the game more
using biofeedback control than with a standard
keyboard/mouse scheme. In our opinion, there are at
least two factors that seem to contribute towards this
outcome.

The first is the novelty factor of the technology: less
than a handful of participants heard about biofeedback
in videogames, and none of them ever had the chance to
try and play videogames with it. The relative success of our
implementation (and even previous works such as~\cite{nacke2011biofeedback, dekker2007please}
can imprint an implicit biasing effect towards higher Fun
ratings. The only way to assess whether the answer would be
different – on a consumer-level setting such as a home use
scenario – is to make a long term study involving several
games (to mitigate game design issues and replicate
accurately the life of the average gamer) with a biofeedback
group and a control group. However, this study would
require a sample population well over 100 subjects for
each group and an extended time frame of over two weeks
or even months. Unfortunately the number of existing
biofeedback games is very little, which makes the idea
infeasible and very hard to execute. The only way to
truly assess the impact of biofeedback is to introduce
it in a commercial context and see how gamers respond
to it and adopt it (or not) over time.

One of our main concerns is that Direct Biofeedback
videogames end up being more exhausting than today's videogames,
due to their strong physical component. Some of our
participants share this concern as well (as shown in
Tables~\ref{tab:keywordresult1}-~\ref{tab:keywordresult2})
and this could negatively impact the average playtime of
game sessions:

\begin{itemize}
\item  \emph{``I think it could become tiring on the long term and it wouldn't
  be so fun (in the situations where you play 10 hours per day)''} (P10)

\item \emph{``During the game it would be cool to switch between unimodal
  and multimodal according to the player’s tiredness''} (P14)

\item  \emph{``It's exhausting after a while...''} (P27)

\item  \emph{``In this game it wouldn't be exhausting because you don't
  run for too long, but if it were mandatory it would be tiring''} (P30 on Biofeedback Sprint)
\end{itemize}

\vspace*{12mm}

On the other hand, some cope well with it and even embrace it
as a necessary consequence of this new gameplay style:

\begin{itemize}
\item  \emph{`It's tiring, but it's a \emph{pleasant} tiring''} (P2 about both Biofeedback versions of Sprint)

\item  \emph{``I also liked to contract the arm in order to activate Use and Grab
  – it was slightly uncomfortable but it was rewarding for the
  achieved realism' (...)'} (P5)

\item  \emph{``In terms of precision pointing a gun and holding breath is
  more tiring, but to appreciate a game it becomes more interesting
  – however on a competitive level it’s less useful''} (P8 on Gun Recoil)
\end{itemize}

\vspace*{12mm}

The second factor that we feel it might impact the Fun ratings is
related with the two Immersion theories described by
Brown~\cite{brown2004grounded} and Ermi~\cite{ermi2007fundamental}.
Brown's ``Engagement'' states that players need to invest time, effort
and attention in order to reach that stage of immersion, once the
game controls and the genre match the player's preferences;
Ermi's ``Challenge-based immersion'' is \emph{``the feeling of immersion
when one is able to achieve a satisfying balance of challenges and
abilities''} and \emph{``can be related to motor skills or mental skills''}~\cite{ermi2007fundamental}.

\begin{itemize}
\item  \emph{``In both biofeedback versions we have to imagine what
  we would normally do to activate the actions, it gives a
  little more of immersion and provides some differences in
  gameplay. It's also more challenging having to find out
  in these two what we have to do.''} (P24)
\end{itemize}

The implicit effort that the biofeedback conditions demanded from players
is a hint that players should feel more immersed in the biofeedback
conditions, and Fun and Enjoyment are indicators that are related
with the broad concept of Immersion.

One participant made an interesting remark in our game about
immersion and biofeedback – we consider that biofeedback can be
positively combined with other technologies to enhance further
the sensation of immersion:

\begin{itemize}
\item  \emph{``In the biofeedback conditions I thought the interactivity
  was quite original [creative], both in the surrounding objects
  (the boxes and handles) as well in the actions that our character
  made like breathing or gun recoil control – they help providing
  the game a bigger feeling of immersion. The real feeling of
  immersion would be to account for head movement, like the Oculus
  Rift\footnote{http://www.oculusvr.com/ - ``The Oculus Rift is a next-generation virtual reality headset designed for immersive gaming.''} device.''} (P11)
\end{itemize}

Many users complained about the lack of an accelerometer (or a similar
device to track spatial orientation) in our GLOVE device because
they wanted to grab objects using only the glove:

\begin{itemize}
\item  \emph{``[in the biofeedback versions] you need to add an accelerometer on the glove''} (P5)

\item  \emph{``The grab mechanic in both conditions does not work well because we
  have to use our hand closed (while grabbing the object) to move the mouse.''} (P8)

\item  \emph{``Grabbing the object and then returning to the mouse... it really
  makes me want to move it with the glove''} (P12)

\item  \emph{``I don't like having to grab and then not being able to use the mouse''} (P21)
\end{itemize}

The mouse solution was not the best in terms of proper game design, but it was
the only one in our reach in terms of available time and human resources.
We hypothesize that Biofeedback can reach its full potential when combined with
other technologies such as orientation trackers (e.g. accelerometer), virtual
reality devices such as the Oculus Rift, or even with the existing motion-devices
for games: the Microsoft Kinect, Nintendo Wii Remote or PlayStation Move.

\section{Unimodal vs. Multimodal Game Design} \label{sec:univsmulti}

Apart from Possess and Underwater, we were not able to uncover
significant differences in Fun between unimodal and multimodal
mechanisms. Players tended to use the ``Realism'', ``Exhaustive'' and
``Complete'' keywords to describe mechanics in the multimodal version
– surprisingly, the unimodal Gun Recoil scored lower in ``Complete''
than the vanilla and multimodal versions. Game controls can be
considered to be slightly harder to use on an increasing scale
from vanilla to unimodal and multimodal – although some unimodal
mechanisms can stand in par with the vanilla version (Gun Recoil,
Fire Blow and Use) according to our statistical results.

Some players reached the consensus that the multimodal was the one
with higher realism, but suffered from unnecessary levels of
exhaustiveness. They valued the simplicity of the unimodal design,
which was less also tiring:

\begin{itemize}
\item  \emph{``Having to contract the arm always to open doors is bad,
  it's better to do that only when firing the gun''} (P1 about the
  multimodal version of the Grab mechanic)

\item   \emph{``In the unimodal version I also thought the experience was
  rewarding because it was quite interactive, simple and fits a
  gamer type that is more rookie and/or lazy – it still forces us
  to put effort but in a 'lighter' way... 'interactivity light'.''} (P5)

\item  \emph{``I think that the multimodal version tries to be more immersive, 
  but it's too complex. The unimodal version can be immersive and simple.
  Both types bring added value when compared to the vanilla version.''} (P10)

\item \emph{``It was confusing having to coordinate both things [the arm
  and breathing] (...) I think it wouldn’t be the best combination
  and having to breath disturbs the action''} (P29 about Multimodal Gun Recoil)
\end{itemize}

\vspace*{12mm}

Others valued the assertiveness and safety of the multimodal design,
which added the highest realism, more sense of control or avoided
sensor faults – either unintended activations of the sensors on
their own (i.e. false positives) or the out-of-context activation
of the players’ abilities:

\begin{itemize}
\item  \emph{``I prefer the complexity of the multimodal version for having
  more details that make the game more realistic.''} (P4)

\item  \emph{``I think it makes sense having distinctions between the
  actions of diving and using a power [Invisibility] (...) the
  unimodal version wasn't very intuitive. (…) The multimodal version
  was more complete for distinguishing torch blowing and Possess''}
  (P10 – Invisibility and Underwater Breathing share the RESP sensor;
  Air Blow and Possess share the TEMP sensor)

\item  \emph{``Closing the hand has more feedback for the game to know
  that it's what I want''} (P12 on Invisibility)

\item  \emph{``I would choose the unimodal version only due to the detail
  of arm contraction in Use and Grab. Disregarding that I would use
  the multimodal version, since it uses ways that are more explicit
  and direct to indicate the actions that we want to perform – Possess,
  Invisibility, running with both legs, etc. (...) Breathing in and
  activating the Invisibility mechanism by accident doesn't seem
  good to me''} (P14)

\item  \emph{``In the multimodal version when I used Possess, I felt
  more control by having to perform a gesture before activating the
  power than by having to blow on the enemies [in the multimodal
  version]. (...) In the unimodal version, having to breathe both
  to stay invisible and to breathe underwater was very distracting.''}
  (P22)
\end{itemize}

Lastly, some participants reported that biofeedback broke game flow
more often, most likely due to the time spent executing physical
actions in comparison to pressing a button. Our ``Playability'' results
in the second questionnaire seem to confirm this as well, but we
think this is a by-product of exposing players to three different
combinations of game controls in such a short time period.
Additionally, players have years and years of familiarity with the
keyboard/mouse control scheme.

\begin{itemize}
\item  \emph{``The multimodal version is a little too complex for a game where
  you have to be fast, with combat and action. The unimodal version
  has the better quantity of interaction when compared to the
  other two.''} (P18)

\item  \emph{``The biofeedback sections don't make the game hard in any way,
  but they make it less fluid in comparison with the vanilla version.
  While in the latter it's possible to walk and activate handles (Item Use)
  quite fast, when using biofeedback you need to stop and use
  the glove.''} (P20)

\item  \emph{``The multimodal version is more complex and breaks a
  little the pace of the game. The unimodal one allows us to
  play in a more fluid way.''} (P26)

\item  \emph{``In the multimodal version, while the actions were more
  realistic, tasks took longer to be executed. In the unimodal mode,
  actions had more realism than the vanilla version and the difference
  in time execution of actions was almost unnoticeable.''} (P29)
\end{itemize}

Based on our players' feedback and our experience, we think that a
mixed approach should be taken instead of a strict implementation
of one of these biofeedback types. Long-term exhaustiveness, ``naturalness''
of the physical action and reliability of mechanism activation
are the guidelines to be followed. As in modern games, books or
movies, it depends on what feelings and sensations the game designer
is trying to convey to the target audience.

For example, doors that open easily should not need overpowering arm
contraction in order to open them: closing the hand and moving the
arm without any contraction is more than enough. Objects that are
light and heavy fall under this category as well. Putting out a
matchstick or a candle should only need the TEMP sensor instead
of breathing in all the air you can hold first. Considering the
panoply of creative ways players can interact with a game, there
may be scenarios where even simple actions should be difficult to
execute. For example, consider a game section where the game
character was shot in a gunfight and barricade him or herself until
reinforcements arrive. However, being heavily injured, the
character does not have much strength left to move the heavy
objects needed to barricade the room. In this scenario it might
be a good option to use multimodal biofeedback to purposely
make this simple interaction more challenging and realistic,
instead of resorting to, for example, quick-time events.

\vspace*{12mm}

Common gameplay situations with harder obstacles (e.g. doors that
will not budge) or more complex actions (e.g. manipulate or
throw objects) can justify the use of Multimodal Biofeedback.
The feeling of empowerment to activate a special power as in
our game (Invisibility, Possess) is something that most
participants felt it worked perfectly. This is another aspect
where the potential of Multimodal Biofeedback is appreciated,
which some of our participants dubbed as the ``hand-sign system'':

\begin{itemize}
\item \emph{``Since I like Naruto it ends up being fun... ''} (P3 about Multimodal Possession)
\item \emph{``I liked the part of making a gesture in Possess to control the enemies.''} (P25)
\item \emph{``In the Possessing power I liked more the multimodal version
        because we needed to make hand signs in comparison with the
        unimodal version where you only had to blow air – the
        multimodal version was more realistic and interactive.''} (P13)
\item \emph{``[except Sprint] In all other mechanisms, I prefer the multimodal version – I liked using a hand symbol combined with another action to activate the mechanisms.''} (P5)
\end{itemize}

\section{Possible Changes in Gameplay} \label{sec:changesgameplay}

The successful integration of Biofeedback gameplay mechanics in videogames
will hardly follow a plug and play paradigm – not at least until a
very mature, standardized period. As such, the design of biofeedback
games is an important facet of their development. Thus, it is
important to examine how our biofeedback game design choices
impacted player experience. Following this thesis, Participant
19 provided the following differential analysis of Biofeedback
and traditional videogames:

\begin{itemize}
\item \emph{``The great difference between games that just use mouse and keyboard and those where sensors are used, for example, is that in games with sensors a higher importance is given to gameplay instead of more properly the game objective – for example in a shooter, by playing with mouse and keyboard the enemies are stronger and in bigger numbers, while when playing with sensors more importance is given to controls and movements of the player''} (P19)
\end{itemize}

Sensors can change the focus and gameplay of the game, which may require certain adaptations on the game logic to fit the use of the new devices. Some of the mechanics in
Tables~\ref{tab:keywordresult1}-~\ref{tab:keywordresult2} were considered slightly more relevant to gameplay (``Core in-game'') in the biofeedback versions.

Regarding Perceived Choice differences in the IMI Questionnaire, Participant 22 said:

\begin{itemize}
\item \emph{``[in biofeedback] Although it was something that was stored in the computer, I felt that the choices inside the game were mine''} (P22)
\end{itemize}

This was the only explicit statement regarding Perceived Choice recorded in the entire study. While participants weren’t very expressive on this topic, the ANOVA tests showed that players experienced a slightly higher sense of Perceived Choice in the biofeedback conditions against the vanilla version of the game, where it was more neutral. It could be because some of the sensors could be activated in more than one way (e.g. EMG-Arm can be activated either by muscle contraction or by simply stretching the arm in any direction), while button-based actions typically are activated always the same way.

The fact that Participant 22 considered that the choices were her own could be because the fact that the game uses the player’s own physiological signals to trigger actions inside the game could create a subconscious bond between the player’s physical and virtual, in-game body. It stops being a matter of manipulating the character inside the game – which does what we tell him/her to do via keyboard/mouse commands – and becomes a situation where players can manifest their will by using their body actively to interact with the game world.

\section{Limitations} \label{sec:limitations}

Our investigation is limited in the sense that we built a game prototype on one genre from all of those that currently exist today (e.g. Racing games, Platformers, etc.). In order to assess the potential of Biofeedback, more prototypes need to be built in other game genres. Nacke et al. also refer this in their previous work and while our work builds on theirs', it is concerned with a vastly different game genre (2D Platformer vs. 3D First-Person Shooter with supernatural abilities).

Secondly, the appropriate hardware, robust frameworks and technology to process the data returned by the physiological devices for Biofeedback games are not yet available as of this time. Calibration of the physiological signals before gameplay was done manually. The fact that players would need to perform calibration manually and equip the biofeedback devices on their own – a process that took us approximately 15-20 minutes – makes Biofeedback infeasible for the everyday consumption of videogames at the current time.

Additionally, some of our participants experienced undesired activations in the TEMP and RESP sensors which we could not solve during the development phase. The TEMP sensor sometimes picked up very small activations that should be considered false positives. The RESP sensor did not have the expected variation due to a very small change in the chest volume while breathing. This mostly happened on participants with a more muscular chest area. 

Everybody has a different physiology: this is an aspect that wasn't reported in other biofeedback games and should be carefully observed in future biofeedback prototypes.

%\begin{quote}
%  \emph{``I wish the diving section was bigger, it was very short.''} (P4)
%\end{quote}

%------------------------------------

%\vspace*{12mm}

\chapter{Future Work} \label{chap:chapter7}

\section*{}

Our initial exploration on this novel type of Direct Biofeedback opens several interesting research avenues for future projects. To the best of our knowledge, our prototype succeeded in combining, for the first time, more than one direct physiological sensor inside one game mechanic. It will be interesting to see how new prototypes build on our finding to further improve user experience and innovate on the design of biofeedback augmented gameplay mechanics. It would also be stimulating to see more theoretical studies on how level design guidelines should be adapted to accommodate for the idiosyncrasies that biofeedback introduces in the game design process. Lastly, it would be even more exciting to see these same studies applied to other game genres (e.g. Adventure, Exploration and Role Playing Games).

One particular detail we noticed is that one of our participants (P8) did not seem as pleased as the other participants regarding the biofeedback interaction. In his words, this type of interaction is not good for those who dedicate themselves to highly competitive game environments (e.g. online games like \emph{``Starcraft II''}, \emph{``Diablo III''} or \emph{``League of Legends''}). This participant believed that this technology was ideal for those who like to appreciate the story or the 3D environment of a game. In our opinion, a study to compare the impact of biofeedback techniques on reported Fun between players with different playing philosophies would potentially yield valuable results. Players could, for example, be segmented as ``narrative appreciator'', ``competitive'', ``social co-operator'' or ``environment fancier''.\footnote{Author's note: at the time of the writing of this document we did not know about the existence of theories on personality and player types, and merely transcribed (literally) the player types suggested by P8. After recently taking knowledge of Bartle's ``Hearts, Clubs, Diamonds, Spades'' player model~\cite{bartle1996hearts}, the four play styles described by Bartle (Killers, Achievers, Explorers, Socializers) show an interesting similarity to the concepts exposed by P8. In conclusion, Bartle's player model and related theories are a possible starting point to explore the connection between play styles and the effectiveness of Biofeedback.}

We were unable to draw meaningful conclusions on whether contextual biofeedback mechanics (e.g. closing a hand has one effect if we are holding a door knob and another when grabbing an enemy by the neck) had any effect on the gameplay experience when compared with other types of biofeedback. In our unimodal prototype, some players affirmed they were confused when using the RESP sensor to hold breath underwater and to activate invisibility outside of the water. TEMP was also used for Blow and Possession and some participants mentioned that it would be good to have a distinction in activation between the two mechanics. According to them, this was achieved in the multimodal version, albeit obviously with an increase in mechanic complexity as previously discussed. Exploring this challenge would be another interesting study from a game design perspective.

On a different note, we are unsure of the reaction players might show towards Biofeedback on a long-term period. The exhaustion factor should be accounted for when designing videogames, as it could become a severe obstacle to the longevity of a game. Even considering that our players spent a considerable amount of time playing the ``GenericShooter 3000'' in both its two biofeedback variants, it is impossible to predict how they are going to adapt. Furthermore, the side effects that repeatedly performing the same bodily actions may produce over extended time periods (years) still need to be thoroughly considered as improper usage may have ill-effects (e.g. carpal tunnel syndrome).

Finally, addressing some of the current limitations in games development for Biofeedback could benefit on-going research. Creating a high level, open-source framework capable of:

\begin{enumerate}
\item Eliminating undesired signal noise, and
\item Performing automatic sensor calibration
\end{enumerate}

would lead to improved usability (less accidental activations) and a reduced dependency on specialized personnel. Addressing these issues would surely potentiate the use of this technology in the near future.

%\begin{quote}
%  \emph{``I wish the diving section was bigger, it was very short.''} (P4)
%\end{quote}

%------------------------------------

%\vspace*{12mm}
%
%Lorem ipsum dolor sit amet, consectetuer adipiscing elit. Phasellus
%tellus pede, auctor ut, tincidunt a, consectetuer in, felis. Mauris
%quis dolor et neque accumsan pellentesque. Donec dui magna,
%scelerisque mattis, sagittis nec, porta quis, nulla. Vivamus quis
%nisl. Etiam vitae nisl in diam vehicula viverra. Sed sollicitudin
%scelerisque est. Nunc dapibus. Sed urna. Nulla gravida. Praesent
%faucibus, risus ac lobortis dignissim, est tortor laoreet mauris,
%dictum pellentesque nunc orci tincidunt tellus. Nullam pulvinar, leo
%sed vestibulum euismod, ante ligula elementum pede, sit amet dapibus
%lacus tortor ac nisl. Morbi libero. Integer sed dolor ac lectus
%commodo iaculis. Donec ut odio.  

\chapter{Conclusions} \label{chap:chapter8}

\section*{}

We introduced the use of Multimodal Direct Biofeedback in videogames and compared it with standard mouse/keyboard control schemes, as well as with traditional Unimodal Direct Biofeedback (as introduced by Nacke~\cite{nacke2011biofeedback}).

While the vanilla version of the game proved itself easier to use than the multimodal version in all mechanisms, not all of the unimodal mechanics were statistically more complex compared to the vanilla version. However, as most participants noted, higher complexity was generally not a negative aspect, as our results show that both biofeedback versions provide a more realistic feel when compared to the vanilla version. We also show that this feeling of heightened realism and depth is higher in the multimodal version – but at the expense of a more exhaustive gaming experience.

The IMI questionnaire also detected a significant contribution of biofeedback towards Fun in gameplay – most likely due to the satisfaction of players seeing their own body movements reflected inside the game. However, using only the IMI questionnaire renders us unable to evaluate the differences in functionality between the unimodal and multimodal types. To compensate for this, we attempted to further study the two biofeedback variants using the remaining questionnaires, the Card Sorting process and our participants' commentaries. These new metric were highly informative and allowed us to conclude that both biofeedback types have their place in the gameplay experience and should be combined simultaneously to add depth to videogames. The major guideline we were able to extract is that they should be combined according to the game's current context (e.g. lifting light objects versus heavy objects).

The recent developments in biofeedback interaction have left us sitting at an interesting crossroad in entertainment mediums: in books, movies and most current games, we assimilate the full gameplay experience while on a comfortable position from which we can abstract ourselves from. However, the latest motion-gaming (e.g. Nintendo Wiimote, PlayStation Move, Microsoft Kinect) are shifting towards more physically-active scenarios. The key to Direct Biofeedback's success could lie on the players' ability to ``step out of the couch'' – and our ability to motivate them to do so -, as they could be no longer ``proxy-experiencing'' the adventures of fictional characters, but rather living their own experiences on a fantasy world.

%%----------------------------------------
%% Final materials
%%----------------------------------------

%% Bibliography
%% Comment the next command if BibTeX file not used
%% bibliography is in ``myrefs.bib''
\PrintBib{myrefs}

%% comment next 2 commands if numbered appendices are not used
\appendix
\chapter{Card Sorting - Plotted Charts} \label{ap1:charts}

For ease of interpretation of the data presented on Tables~\ref{tab:keywordresult1}-~\ref{tab:keywordresult2}, we plotted the results of the card sorting process. The results are grouped in the same manner they are presented on Tables~\ref{tab:keywordresult1}-~\ref{tab:keywordresult2}:

\begin{itemize}
\item Useless in-game / Core in-game;
\item Incomplete / Complete;
\item Imaginary / Realistic;
\item Relaxed / Exhausting;
\item Simple / Complex;
\item Confusing / Intuitive.
\end{itemize}

Notice, however, that the presented comparison pairs are merely one way to
interpret the collected data. Other pairs could be established to
extract useful information - for example: ``Complex vs. Confusing'',
``Simple vs. Intuitive'' or ``Realistic vs. Exhausting''.

\begin{sidewaysfigure}[t]
  \begin{center}
    \leavevmode
    \includegraphics[width=0.85\textwidth]{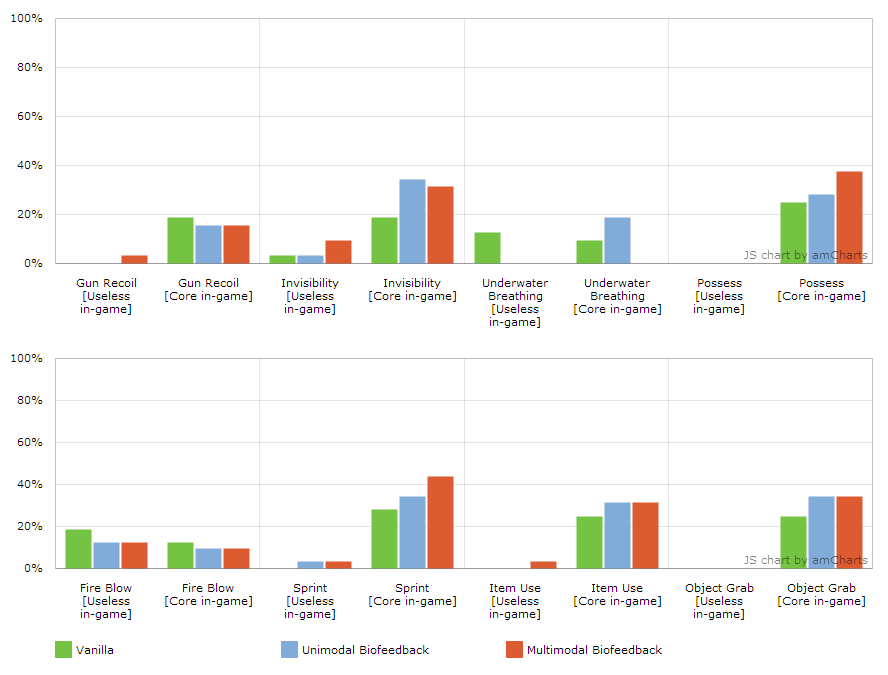}
    \caption[Card Sorting Chart - Useless in-game vs. Core in-game]{Useless in-game vs. Core in-game.}
    \label{fig:useless-core}
  \end{center}
\end{sidewaysfigure}

\begin{sidewaysfigure}[t]
  \begin{center}
    \leavevmode
    \includegraphics[width=0.85\textwidth]{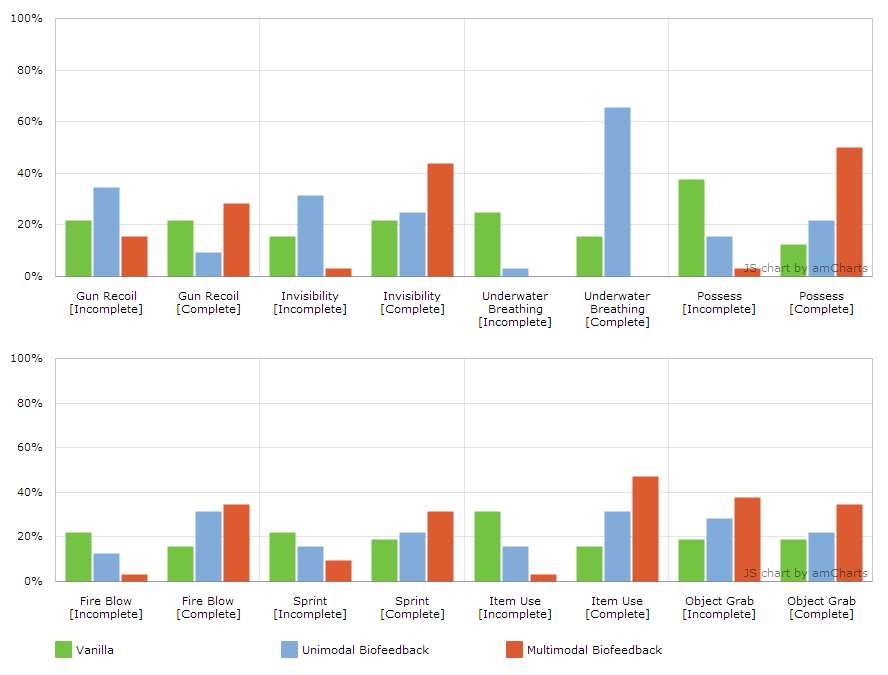}
    \caption[Card Sorting Chart - Incomplete vs. Complete]{Incomplete vs. Complete.}
    \label{fig:incomplete-complete}
  \end{center}
\end{sidewaysfigure}

\begin{sidewaysfigure}[t]
  \begin{center}
    \leavevmode
    \includegraphics[width=0.85\textwidth]{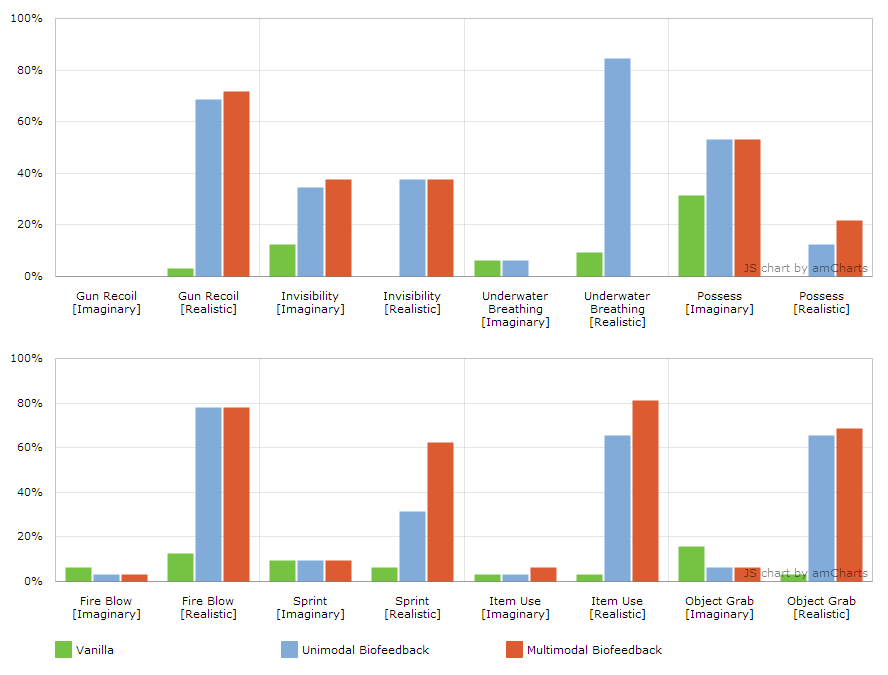}
    \caption[Card Sorting Chart - Imaginary vs. Realistic]{Imaginary vs. Realistic.}
    \label{fig:imaginary-realistic}
  \end{center}
\end{sidewaysfigure}

\begin{sidewaysfigure}[t]
  \begin{center}
    \leavevmode
    \includegraphics[width=0.85\textwidth]{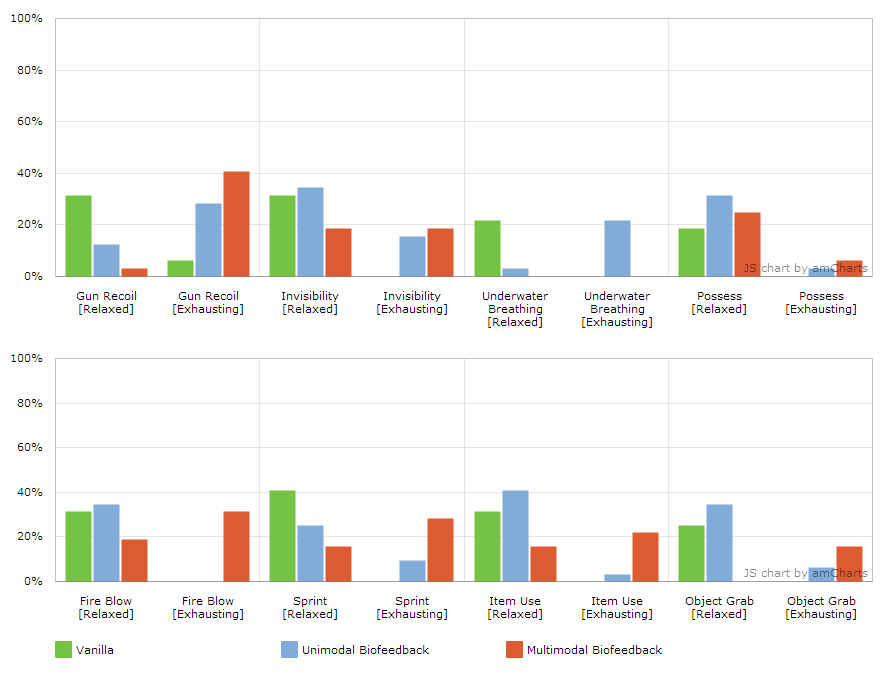}
    \caption[Card Sorting Chart - Relaxed vs. Exhausting]{Relaxed vs. Exhausting.}
    \label{fig:relaxed-exhausting}
  \end{center}
\end{sidewaysfigure}

\begin{sidewaysfigure}[t]
  \begin{center}
    \leavevmode
    \includegraphics[width=0.85\textwidth]{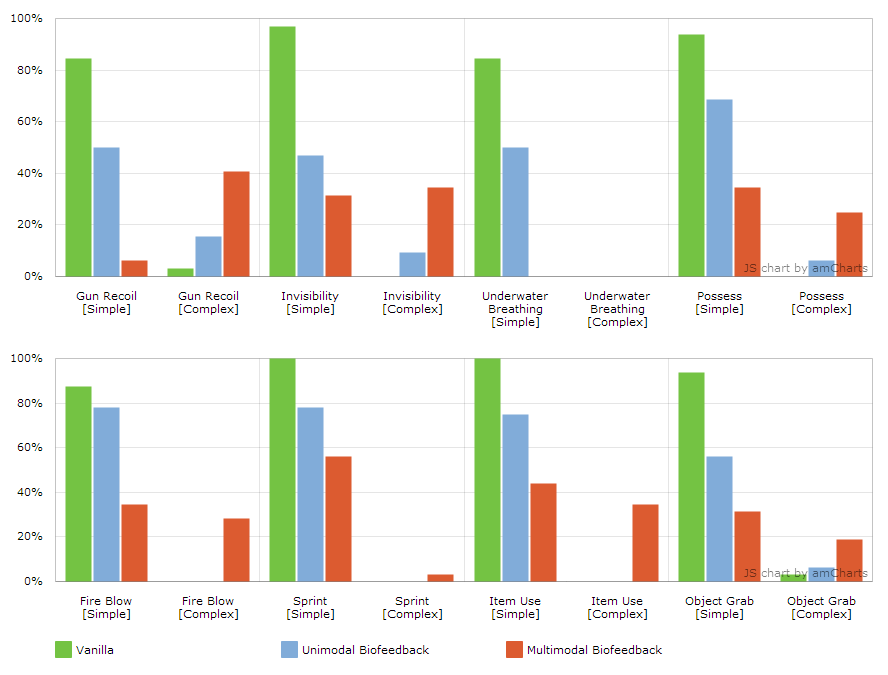}
    \caption[Card Sorting Chart - Simple vs. Complex]{Simple vs. Complex.}
    \label{fig:simple-complex}
  \end{center}
\end{sidewaysfigure}

\begin{sidewaysfigure}[t]
  \begin{center}
    \leavevmode
    \includegraphics[width=0.85\textwidth]{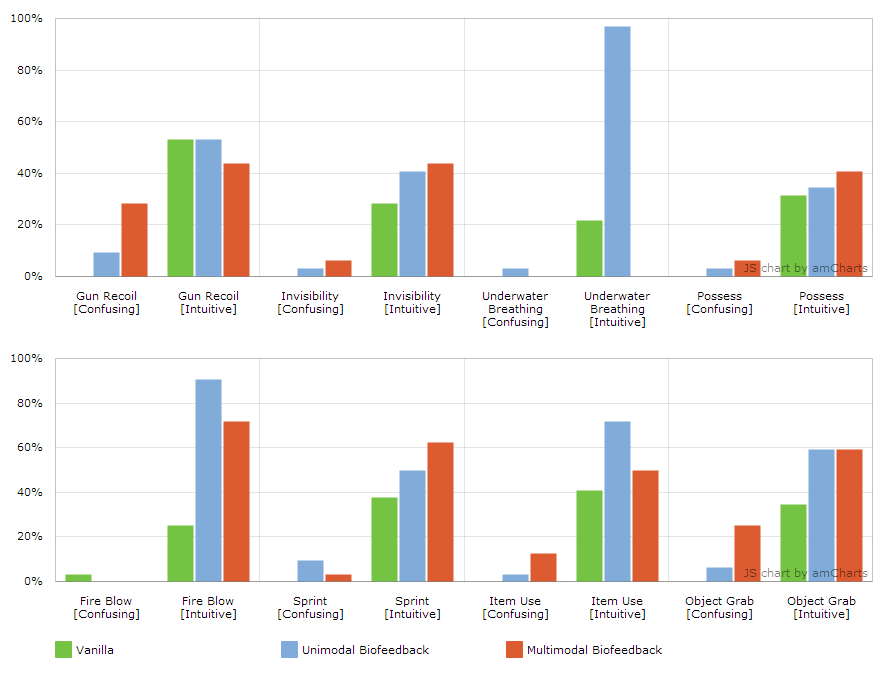}
    \caption[Card Sorting Chart - Confusing vs. Intuitive]{Confusing vs. Intuitive.}
    \label{fig:confusing-intuitive}
  \end{center}
\end{sidewaysfigure}

%% Index
%% Uncomment next command if index is required
%% don't forget to run ``makeindex mieic-en'' command
%\PrintIndex

\end{document}